%% file: main.tex
\documentclass[11pt,a4paper]{article}
\pdfoutput=1 

\newcommand{\fixme}[1]{}

\newcommand{\mus}{\ensuremath{\muup}s\xspace}
\newcommand{\utca}{MicroTCA\xspace}

\usepackage{jinstpub}
\usepackage{algorithmic}
\usepackage{xspace}
\usepackage{xfrac}
\usepackage{newtxmath}
\usepackage[tight]{units}
\usepackage{nicefrac}
\usepackage{lineno}


\author[a]{E.~Bartz,}
\author[b]{G.~Boudoul,}
\author[c]{R.~Bucci,} 
\author[d,1]{J.~Chaves, 
\note{Currently at Johnson and Johnson, New Brunswick, NJ,
  USA.}}
\author[e]{E.~Clement,}
\author[d]{D.~Cranshaw,}
\author[f]{S.~Dutta,}
\author[a]{Y.~Gershtein,}
\author[g]{R.~Glein,}
\author[h]{K.~Hahn,}
\author[a]{E.~Halkiadakis,}
\author[c]{M.~Hildreth,}
\author[a]{S.~Kyriacou,} 
\author[c]{K.~Lannon,}
\author[i]{A.~Lefeld,}
\author[h]{Y.~Liu,}
\author[g]{E.~MacDonald,}
\author[j]{N.~Pozzobon,}
\author[d]{A.~Ryd,}
\author[c]{K.~Salyer,}
\author[c]{P.~Shields,}
\author[k]{L.~Skinnari,}
\author[g]{K.~Stenson,}
\author[a]{R.~Stone,}
\author[d]{C.~Strohman,}
\author[d]{K.~Sung,}
\author[d,2]{Z.~Tao,\note{Currently at University of British Columbia, Vancouver BC, Canada.}}
\author[h,3]{M.~Trovato,\note{Currently at Argonne National Laboratory, Argonne, IL.}}
\author[g]{K.~Ulmer,}
\author[b]{S.~Viret,}
\author[i]{B.~Winer,}
\author[d,4]{P.~Wittich\note{Corresponding author.},}
\author[i]{B.~Yates,}
\author[d]{and M.~Zientek.}

\emailAdd{wittich@cornell.edu}

\affiliation[a]{Rutgers University, New Brunswick, NJ, USA}
\affiliation[b]{Universit\'e de Lyon, Universit\'e Claude Bernard Lyon 1, CNRS/IN2P3, IP2I Lyon, UMR 5822, Villeurbanne, France}
\affiliation[c]{University of Notre Dame, South Bend, IN, USA}
\affiliation[d]{Cornell University, Ithaca, NY, USA}
\affiliation[e]{University of Bristol, Bristol, UK}
\affiliation[f]{Saha Institute of Nuclear Physics, Kolkata, India}
\affiliation[g]{University of Colorado, Boulder, CO, USA}
\affiliation[h]{Northwestern University, Evanston, IL, USA}
\affiliation[i]{The Ohio State University, Columbus, OH, USA}
\affiliation[j]{Universit\`{a} degli Studi di Padova and INFN Sezione di Padova, Pavoda, Italy}
\affiliation[k]{Northeastern University, Boston, MA, USA}

\arxivnumber{1910.09970} 

\title{
  FPGA-based tracking for the CMS Level-1 trigger using the tracklet algorithm
}

\abstract{
The high instantaneous luminosities expected following the upgrade of the Large Hadron Collider (LHC) to the High-Luminosity LHC (HL-LHC)
  pose major experimental challenges for the CMS experiment.  
  A central component to allow efficient operation under these conditions is the reconstruction of charged
  particle trajectories and their inclusion in the hardware-based trigger system.  
  There are many challenges involved in achieving this: a large
  input data rate of about \unit[20--40]{Tb/s}; processing a new batch of input data
  every \unit[25]{ns}, each consisting of about 15,000 precise position measurements
  and rough transverse momentum measurements of particles (``stubs''); performing the
  pattern recognition on these stubs to find the trajectories; and producing the
  list of trajectory parameters within \unit[4]{\mus}.  
  This paper describes a proposed solution to this problem, specifically, it presents a novel approach to pattern recognition and charged particle trajectory reconstruction using an all-FPGA solution.   
  The results of an end-to-end demonstrator system, based on Xilinx Virtex-7 FPGAs, that
  meets timing and performance requirements are presented along with a further improved, optimized version of the algorithm 
  together with its corresponding expected performance.}

\def\pt{\ensuremath{p_{\mathrm{T}}}\xspace}

\def\ttbar{\ensuremath{\mathrm {t \bar{t}}}\xspace}
\def\pp{\ensuremath{\mathrm{pp}}\xspace}

\begin{document}

\maketitle
\flushbottom

\input{intro}
\input{algo}
\input{firmware}
\input{demonstrator_future}
\input{hardware_impl}
\input{performance}

\section{Conclusions}
\label{sec:conclusions}
For the High-Luminosity LHC upgrade, the CMS experiment will require a new tracking
system that enables the identification of charged particle
trajectories in real-time to maintain high efficiencies for
identifying physics objects at manageable rates.  The tracklet
approach is one proposed implementation of the real-time
track finding. The method is based on a road-search algorithm and uses
commercially available FPGA technology for maximum flexibility.  An
end-to-end system demonstrator consisting of a slice of the detector
in azimuth has been implemented using a Virtex-7 FPGA-based \utca
blade. The final system, which is to be deployed in 2025, will use
future-generation FPGAs.  To scale the demonstrator to the final
system, only a small extrapolation is required. Currently, the
demonstrator only covers the $+z$ side of the detector; in the full
system, both sides will be covered. The detector is largely symmetric
in $\pm z$, so the addition of the $-z$ side only results in increased
occupancy, which is handled by more instances of already-existing HDL
modules.
Since more data are coming into the sector processor, about a factor
of two, the
total I/O requirements will increase by roughly a factor of three,
taking into account both the increase of the total data rate and the
cabling scheme of the new detector. These I/O requirements are within
the capabilities of the specifications of the Xilinx Virtex UltraScale+ 
family of FPGAs. None of these
changes represent more than an evolution of the demonstrator.  The
demonstrator  has been used to validate the algorithm and
board-to-board communication, to measure timing and latency, and to
establish the algorithm performance.  Studies from the demonstrator,
processing events from the input stubs to the final output tracks,
show that the tracklet algorithm meets timing and efficiency
requirements for the final system. 

The tracklet 1.0 demonstrator showed the feasibility of the tracklet 
approach for L1 tracking. Following this demonstration a number
of algorithmic improvements were developed and described in this paper
for tracklet 2.0. 
The L1 tracking algorithm was shown to achieve high efficiencies, e.g. an efficiency of 90\% or higher for charged particles in \ttbar events, as well as precise track parameter resolutions, such as a \unit[1]{mm} (\unit[5]{mm}) $z_0$ resolution at central (forward) $|\eta|$, and a relative $\pt$ resolution ranging between 1--4\%. This performance is more than sufficient for how the tracks will be used in the downstream L1 trigger. 

Beyond tracklet 2.0, improvements are considered
that will enhance the efficiency for displaced tracking as well as reduce the
resource needs and latency by combining different processing steps.
In particular, we can combine the ProjectionRouter, MatchEngine,
and MatchCalculator modules into one combined module and
the TrackletEngine and the TrackletCalculator into another module.
A combined ProjectionRouter, MatchEngine, and MatchCalculator 
module, 'MatchProcessor', would reduce the need for storing the 
projections and candidate matches in memories between the processing 
steps. At the same, it would reduce the latency by only incurring the processing time of one
processing step, but with a slightly longer step latency.

\acknowledgments

 The authors would like to thank the Wisconsin CMS group for their
 support of the CTP7 platform.  We are grateful to the CMS Collaboration for use of their detector simulation software. This work was supported by the U.S.~National Science Foundation through the grants
 NSF-PHY-1607096,
 NSF PHY-1312842,
 NSF-PHY-1307256,
 NSF-PHY-1120138
 and the U.S.~Department of Energy Office of Science DE-SC0011726 
and DE-SC0015973.

\bibliographystyle{JHEP}
\bibliography{main}

\end{document}

%% file: intro.tex
\section{Introduction}

This paper describes a novel implementation of a charged particle trajectory reconstruction approach based on Field-Programmable Gate Arrays (FPGAs) for the CMS experiment~\cite{cms} at the CERN LHC.  The LHC accelerator complex will undergo major upgrades, to be completed in 2026, to increase the instantaneous luminosity to approximately $\unit[7.5 \times 10^{34}]{\mathrm{cm}^{-2}\mathrm{s}^{-1}}$~\cite{hllhc}. The ``High-Luminosity LHC'' (HL-LHC) upgrades will enable searches for undiscovered rare particle physics processes as well as detailed measurements of the properties of the Higgs boson.  The HL-LHC will collide proton bunches every \unit[25]{ns}, and each of these bunch collisions (an ``event'') will consist of an average of 200 proton-proton (\pp) collisions. Only a small fraction of these collisions are of interest for further studies. A fast real-time selection, referred to as the Level-1 (L1) ``trigger'', is applied to decide whether a given collision should be considered for further analysis. The L1 trigger is implemented in custom hardware.  The number of overlapping \pp collisions per event, referred to as pileup (PU), in the HL-LHC era represents a large increase over previous data-taking eras ($\sim$200 at the HL-LHC versus $\sim$30 during LHC Run-2), resulting in a significant challenge to the CMS trigger system -- new handles are required. One such handle is the inclusion of information from the charged particle tracking system. 
This will be the first time that information from a solid-state tracking detector has been included in the hardware trigger at the high collision rates of the LHC.

\begin{figure}
\begin{center}
\includegraphics[width=7cm]{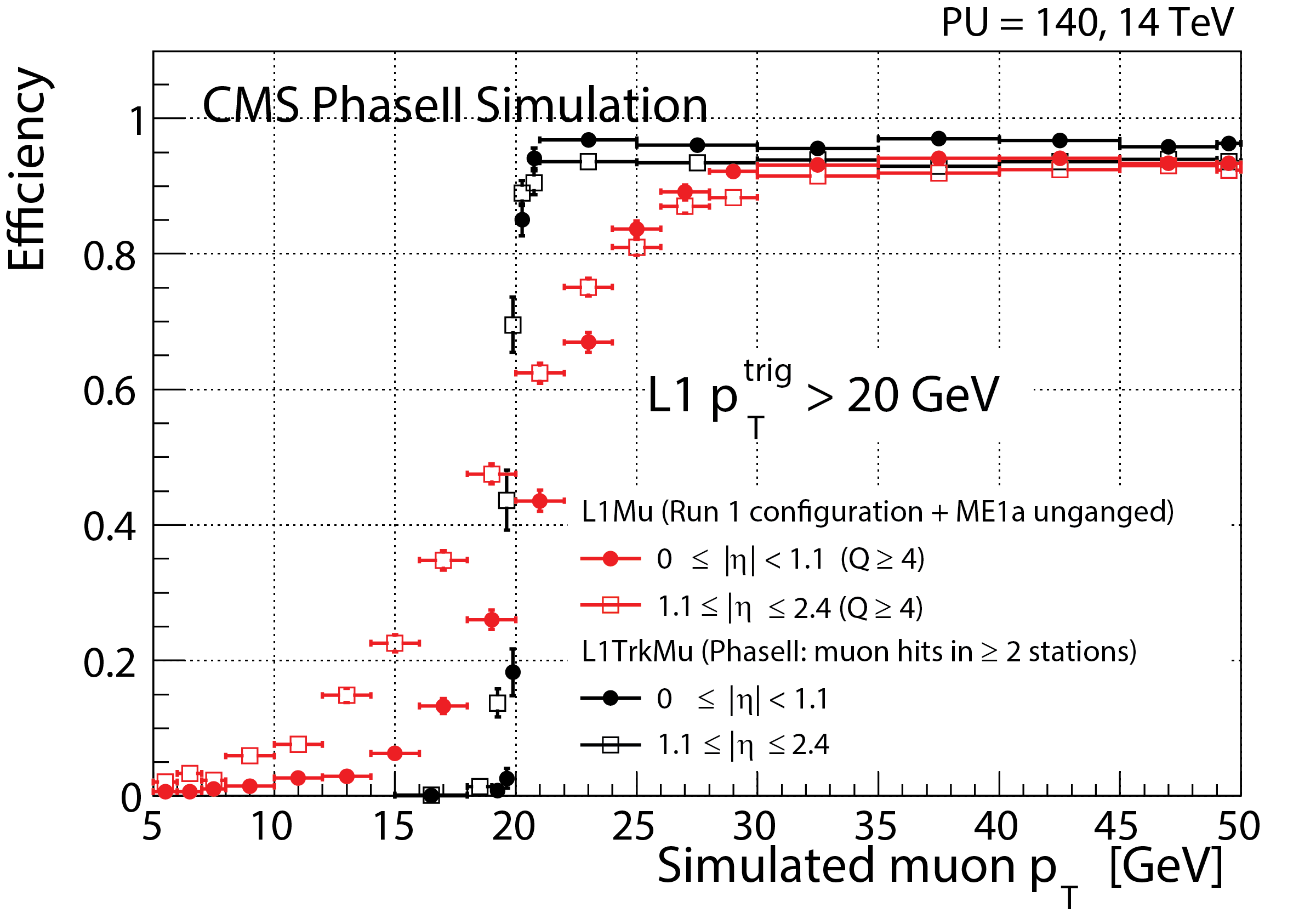}
\includegraphics[width=7cm]{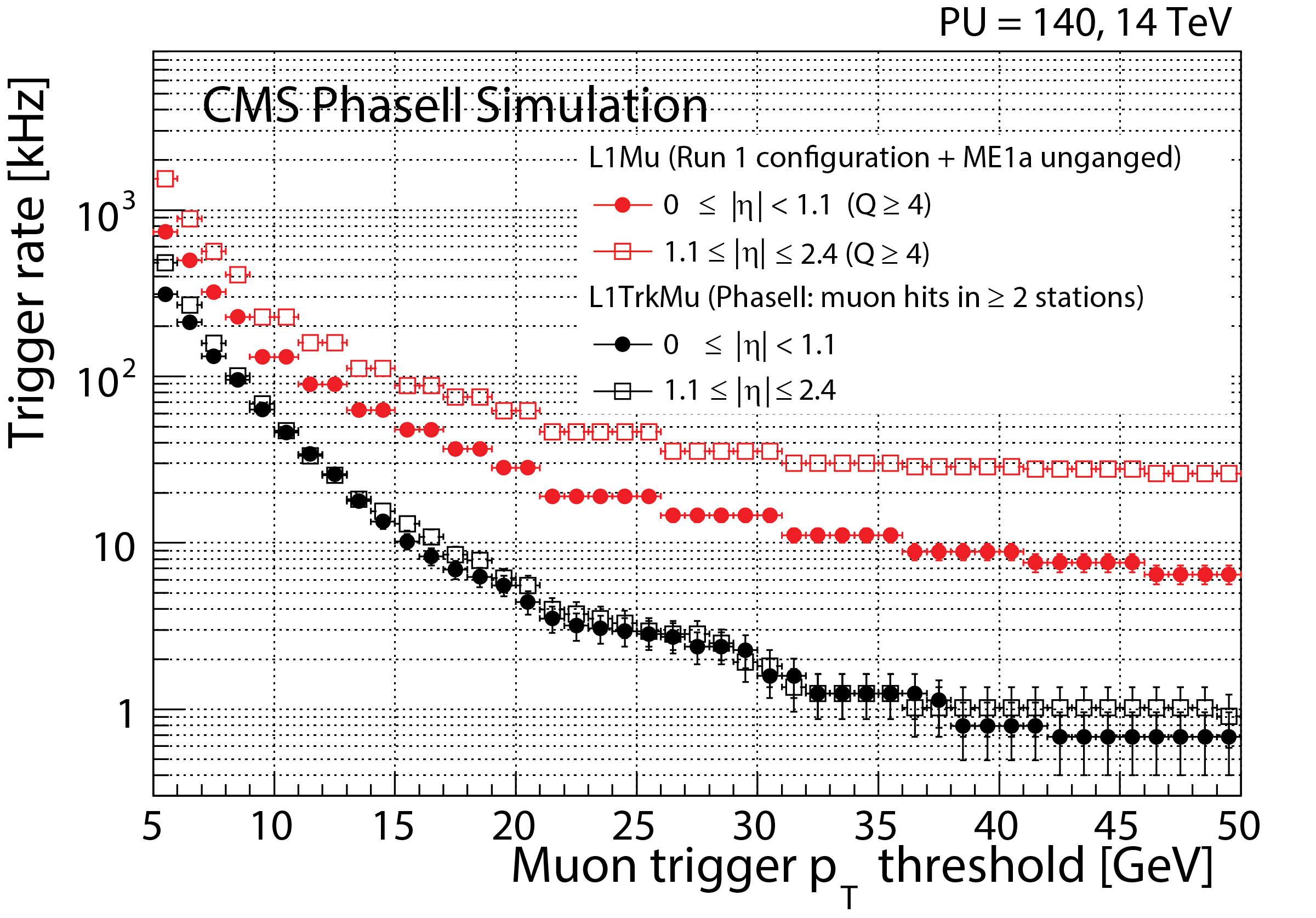}
\caption{The efficiency of a single muon trigger with a L1 threshold of $\pt> \unit[20]{GeV}$
  as a function of muon \pt (left) and the trigger rate as a function of the
  muon trigger threshold (right), shown for the stand-alone muon
  trigger (red) and when including L1 tracking (black) for various
  pseudorapidity ($\eta$) ranges~\cite{tp}.}
\label{fig:HL-LHC_L1}
\end{center}
\end{figure}

Integrating charged particle tracking in the L1 trigger will improve lepton identification and momentum measurements as well as provide track isolation and vertex reconstruction. These additional handles have the potential to reduce the L1 trigger rates while maintaining trigger thresholds acceptable for the CMS physics program.  One example of this is shown in Figure~\ref{fig:HL-LHC_L1}, where the efficiency of a single muon trigger as a function of the muon generated $\pt$ (left) and the corresponding L1 trigger rate (right) are shown~\cite{tp}.  The red curves show the behavior of a stand-alone L1 muon trigger system, while the black curves show the performance of the same triggers when including L1 tracking.  In particular, L1 tracking improves the momentum measurement, which translates to a sharper turn-on curve at the trigger threshold and hence a reduced trigger rate. The tracks must be delivered within \unit[4]{\mus} in order to be used in the trigger decision.

\begin{figure}
  \begin{center}
    \includegraphics[width=0.694\textwidth]{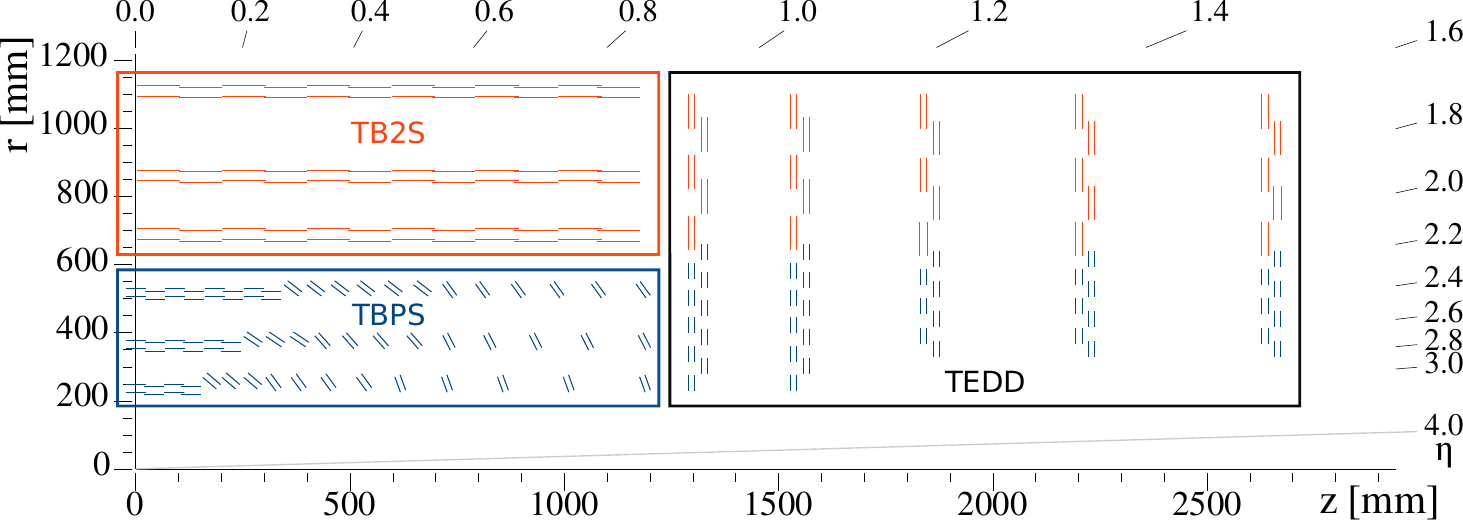}\hspace{2mm}
    \includegraphics[width=0.28\textwidth]{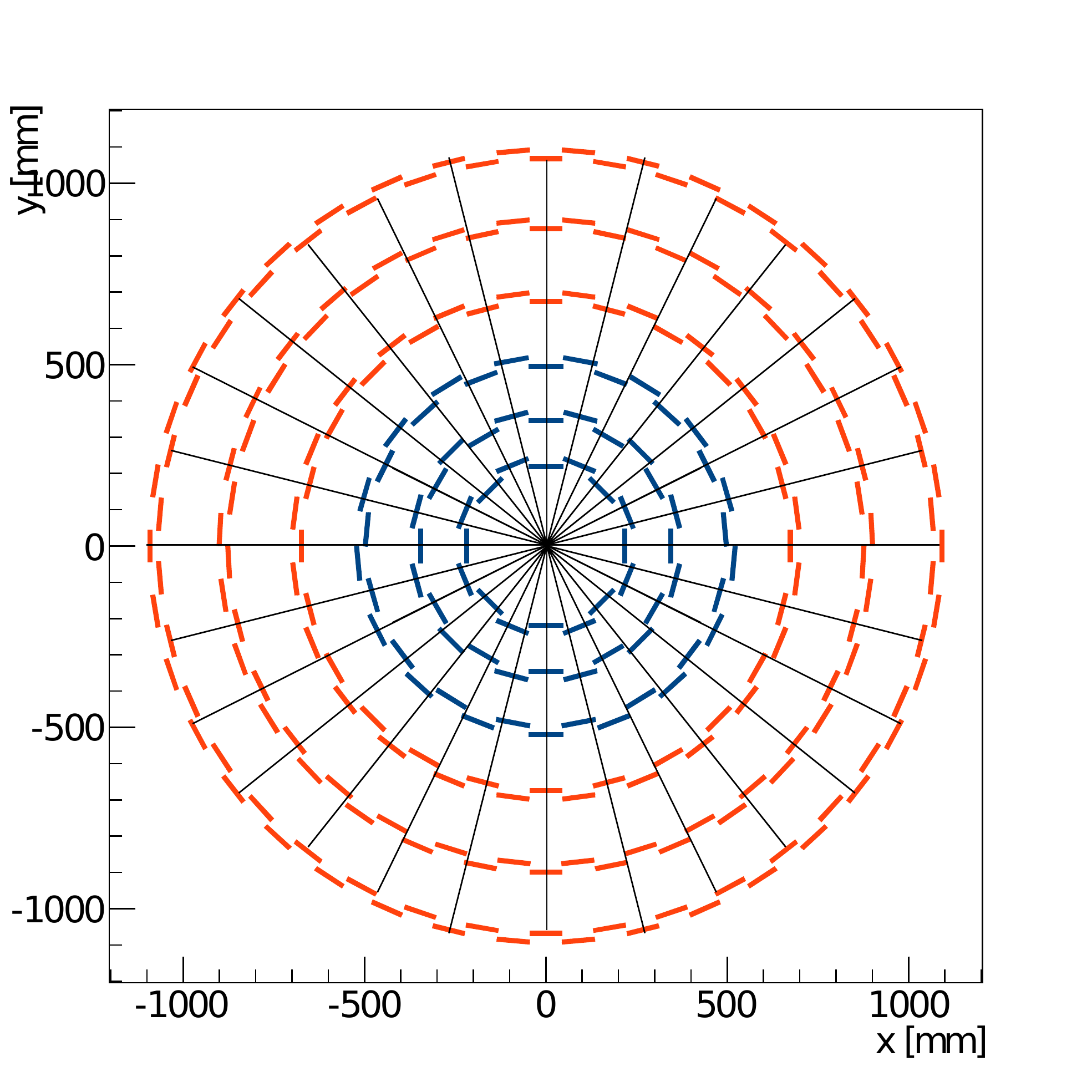}
    \caption{\label{fig:detlayout} One quarter of the layout of the
      upgraded CMS charged particle tracking detector (left), as
      proposed in Ref.~\cite{tracker_tdr}, showing the Tracker Barrel
      with Pixel Strip modules (TBPS) or Strip-Strip modules (TB2S)
      and the Tracker Endcap Double-Disks (TEDD). Not shown is the
      inner pixel part of the CMS charged particle tracking detector,
      as it is not available to use for the L1 tracking.  The LHC
      beams cross at (0,0). In the central barrel region (extending to
      $|z| \lesssim \unit[125]{\mathrm{cm}}$), there are six layers at
      radii from \unit[23]{cm} to \unit[110]{cm}. In the forward
      region (covering $|z| \gtrsim \unit[125]{\mathrm{cm}}$), there
      are five disks at $z$ positions from \unit[130]{cm} to
      \unit[270]{cm}. The detector is divided into \emph{sectors},
      slices in $\phi$ that cover the entire length of the detector
      along $z$, for L1 track finding.  In the $x$-$y$ view of the
      barrel (right), 28 sectors are shown.}
  \end{center}
\end{figure}

To reconstruct the trajectories of charged particles, the CMS experiment includes a tracking detector, which will be replaced for the HL-LHC operation. 
This upgraded tracking detector will consist of an inner tracker based on silicon pixel modules (not available in the L1 trigger), and an outer tracker based on silicon modules with strips and macro-pixel sensors~\cite{tracker_tdr}.
The layout of the upgraded outer tracker, as proposed in Ref.~\cite{tracker_tdr}, is illustrated in Figure~\ref{fig:detlayout}\footnote{
We use a right-handed coordinate system, with the origin at the nominal interaction point, the $x$ axis pointing to the center of the LHC, the $y$ axis pointing up (perpendicular to the LHC plane), and the $z$ axis along the counterclockwise-beam. The azimuthal angle $\phi$ is measured in the $x$-$y$ plane.}. 
Particles are produced at the interaction point and travel outwards in a \unit[3.8]{T} uniform magnetic field, which is parallel to the $z$ axis. The trajectory of a charged particle traversing this magnetic field is bent such that it forms a helix. In the $r$-$\phi$ plane, the helix forms a circle. The radius of this circle is proportional to the momentum in this plane, the transverse momentum, or $\pt$, of the particle.
The tracking detector is based on the concept that charged particles leave energy deposits (``hits'') when crossing the sensitive detector material.  
In the upgraded outer tracker, closely spaced pairs of such hits will be linked on the detector front-end to form ``stubs''. With a sensor spacing of \unit[1--4]{mm}, the relative position of the pairs of hits can be used to read out only those stubs that are likely to come from a particle with $\pt>\unit[2]{GeV}$ (corresponding to a radius of curvature greater than \unit[1.75]{m}). This momentum selection on the stubs reduces the readout bandwidth requirement by a factor of 10~\cite{tp}.  
In the L1 trigger, the stubs can be linked together to reconstruct the trajectories of the charged particles.  

The upgraded outer tracker consists of a central portion with six detector layers parallel to the beam line, called the barrel, and five layers perpendicular to the beam line at large $|z|$, called the disks.  
The inner three layers in the barrel consist of so-called Pixel-Strip (PS) modules. 
The modules have a top sensor with \unit[2.5]{cm} long silicon strips with \unit[100]{$\muup$m} ``pitch'' (segmentation in the bending plane), and a bottom sensor consisting of \unit[1.5]{mm} wide macro-pixels, again with \unit[100]{$\muup$m} pitch.
The inner three layers of the barrel are further divided into two regions: the ``flat'' region, near $z=0$, where the modules are parallel to the beamline, and the ``tilted'' region, at higher $|z|$, where the modules are tilted toward the interaction point. 
The outer three barrel layers instead use ``Strip-Strip'' (2S) modules, where both sensors have \unit[5]{cm} long strips with \unit[90]{$\muup$m} pitch.
The modules on each of the five disks in each half of the detector are arranged in concentric circles, or ``rings'', around the beamline. The two disks closest to the interaction point in each half of the detector have $15$ rings (ten consisting of PS modules and five of 2S modules), and the remaining three have twelve rings (seven consisting of PS modules and five of 2S modules).
Both PS and 2S modules provide a precise position and momentum coordinate for stubs in the transverse plane, while only the PS modules give a precise measurement of the $z$ coordinate. 
Further details on the upgraded CMS tracker detector are available in Ref.~\cite{tracker_tdr}.

This paper discusses the linking of the stubs to form trajectories of 
charged particles, known as L1 tracking, and its associated challenges. 
In each collision of counter-rotating proton bunches (every \unit[25]{ns}), 
about 15,000 stubs are formed.  However, only about 10\% of the
stubs belong to trajectories of interest, so the majority of stubs needs to be
filtered. 
The remaining stubs are combined to form, on average, 180
trajectories every \unit[25]{ns}.  This is the first time that data from the
tracking detector is included in the CMS L1 trigger; previously, the
amount of data to be processed and the computational complexity 
required, placed this task out of reach of FPGAs.

To summarize, some of the challenges involved in reconstructing trajectories of charged particles for the CMS L1 trigger are:
\begin{itemize}
\item Absorb approximately 15,000 stubs arriving each \unit[25]{ns}. The total estimated input bandwidth is in the range of \unit[20--40]{Tb/s}, depending on the beam conditions and trigger conditions.
\item Perform pattern recognition to identify the stubs that belong to
  a given trajectory, rejecting stubs from low-momentum particles.
\item Fit the stubs to extract optimal trajectory parameters.
\item Complete all above steps within the available processing time (``latency'') of \unit[4]{\mus}, in order to feed into the
  decision of whether to retain the event, or discard it,
  before the on-detector data buffers are exhausted.
\end{itemize}

The ``tracklet'' approach for real-time track reconstruction in the hardware-based trigger system of CMS, presented in this paper, is one of three possible implementations that were considered by the collaboration. The other two approaches that were studied were a Hough-transform based approach using FPGAs~\cite{tmtt_demo_paper} and an associative memory based approach using a custom ASIC~\cite{amfpga_fedi_ctd2016}.  The tracklet approach is a ``road-search'' algorithm, implemented using commercially available FPGA technology.  Ever-increasing capability and programming flexibility make FPGAs ideal for performing fast track finding.  The tracklet approach allows a naturally pipelined implementation with a modest overall system size. It also allows for simple software emulation of the algorithm.  We present here the design of such a system and results from a hardware demonstrator system that implements end-to-end reconstruction, from input stubs to output trajectories, within the available latency and with a reasonable system size, for a slice of the detector. In addition to the results from the hardware demonstrator, some further improvements to the algorithm are presented.

Many software-based particle tracking algorithms use a road-search technique where track seeds are found and the trajectories extrapolated to look for matching stubs. This technique works well with the high-precision hits in particle detectors such as the CMS tracker. In the barrel part of the detector, the typical spatial position resolution of the stubs is about \unit[30]{$\muup$m} in $r$--$\phi$ and either \unit[0.5]{mm} (inner layers) or \unit[1.5]{cm} (outer layers) in $z$ in a cylindrical detector volume of about \unit[2]{m} in diameter and \unit[5]{m} in length~\cite{tracker_tdr}. Therefore, the search window (road) around the projected trajectory is small and the probability of finding false matches is low. However, with previous generations of FPGAs, the required computational power for implementing this type of tracking algorithm in the trigger was not available.  Today, the large number of Digital Signal Processing (DSP) blocks and Random Access Memory (RAM) resources available in FPGAs make such an approach feasible. The use of FPGAs for the implementation of this algorithm provides a good match to the requirements. FPGAs provide high-speed serial, low-latency links that are well suited to bring the data into the FPGA for processing. The resources of FPGAs allow the implementation of a highly parallelized architecture, and their reconfigurability allows flexibility to changing needs. The precise space points from the stubs are used to determine trajectories of the particles using the DSP blocks in the FPGA. The DSP blocks are also used to calculate the final track parameters. The RAMs in the FPGA are used to implement the data movement and distributed storage required for this highly pipelined algorithm.


This paper is organized as follows. First, the tracklet algorithm is discussed in Section~\ref{sec:algo}.  Two configurations of the system are discussed in this paper. The first configuration (referred to as ``tracklet 1.0'') corresponds to that employed for a test that implemented a vertical slice of the system as a hardware demonstrator. Since the hardware demonstrator, extensions to the tracklet approach have been implemented that further improve the load balancing and resulting physics performance. This corresponds to the second configuration (referred to as ``tracklet 2.0'').  Except as indicated, this Section describes both tracklet 1.0 and tracklet 2.0.  Section~\ref{sec:firmware_impl} explains the structure of the firmware, which is common for both configurations. In Section~\ref{sec:demonstrator} we detail the results of the hardware demonstrator test, using the tracklet 1.0 configuration.  Section~\ref{sec:hw_platform} describes the foreseen overall system architecture.  Section~\ref{sec:performance} reports on the physics performance of the current system (tracklet 2.0).  Further developments are foreseen. These are discussed in Section~\ref{sec:conclusions}, along with conclusions.



%% file: algo.tex
\section{Tracklet algorithm}
\label{sec:algo}
The goal of the real-time hardware-based track finding is to reconstruct the charged particle trajectories as an estimator of their momenta for particles with $\pt > \unit[2]{GeV}$, and to identify the track $z_0$ position 
with about \unit[1]{mm} precision. The $z_0$ resolution is similar to the expected average separation of proton--proton collisions in the bunch collisions of the upgraded LHC, and thus allows precise determination of the collision vertices.  The proposed tracklet method forms track seeds, ``tracklets'', from pairs of stubs in adjacent layers or disks.  The tracklets provide roads to search for compatible stubs that are attached to form track candidates. A linearized $\chi^2$-fit determines the final track parameters.

The tracklet algorithm has been optimized to provide full coverage of the tracker with a small amount of redundancy in data duplication.  

\subsection{Algorithm overview}

\begin{figure*}
\begin{center}
\includegraphics[width=0.30\textwidth]{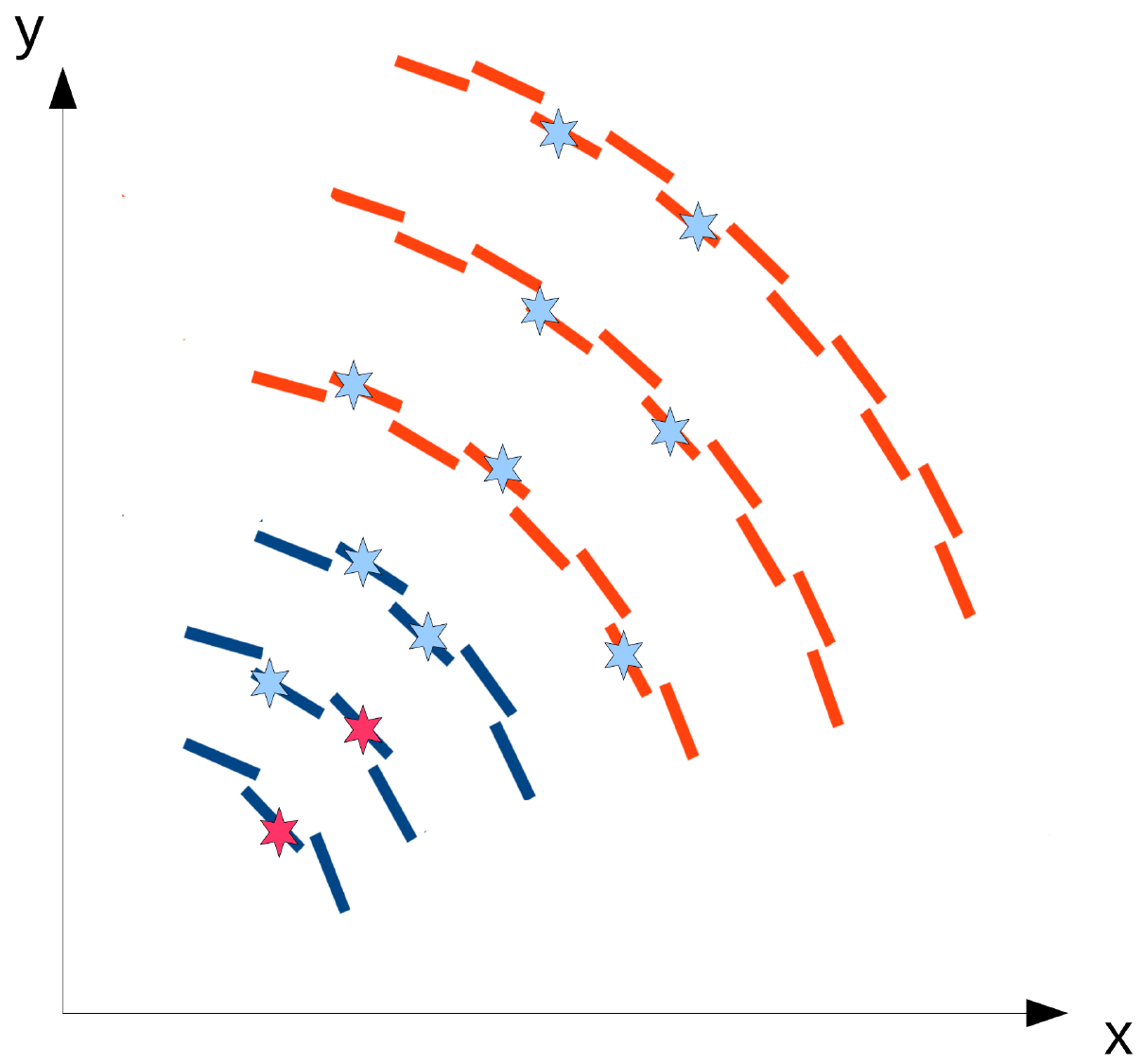}
\includegraphics[width=0.30\textwidth]{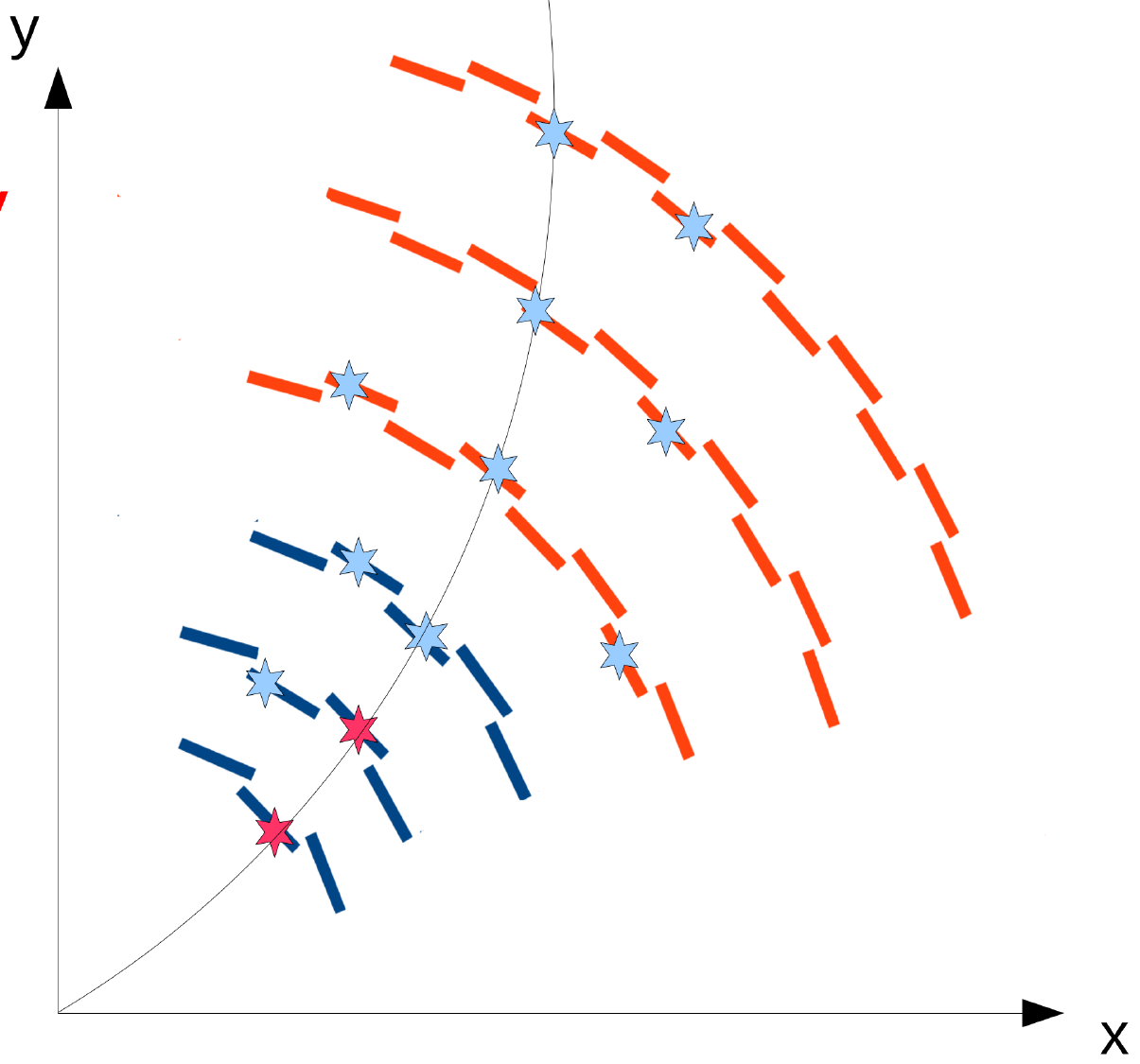}
\includegraphics[width=0.30\textwidth]{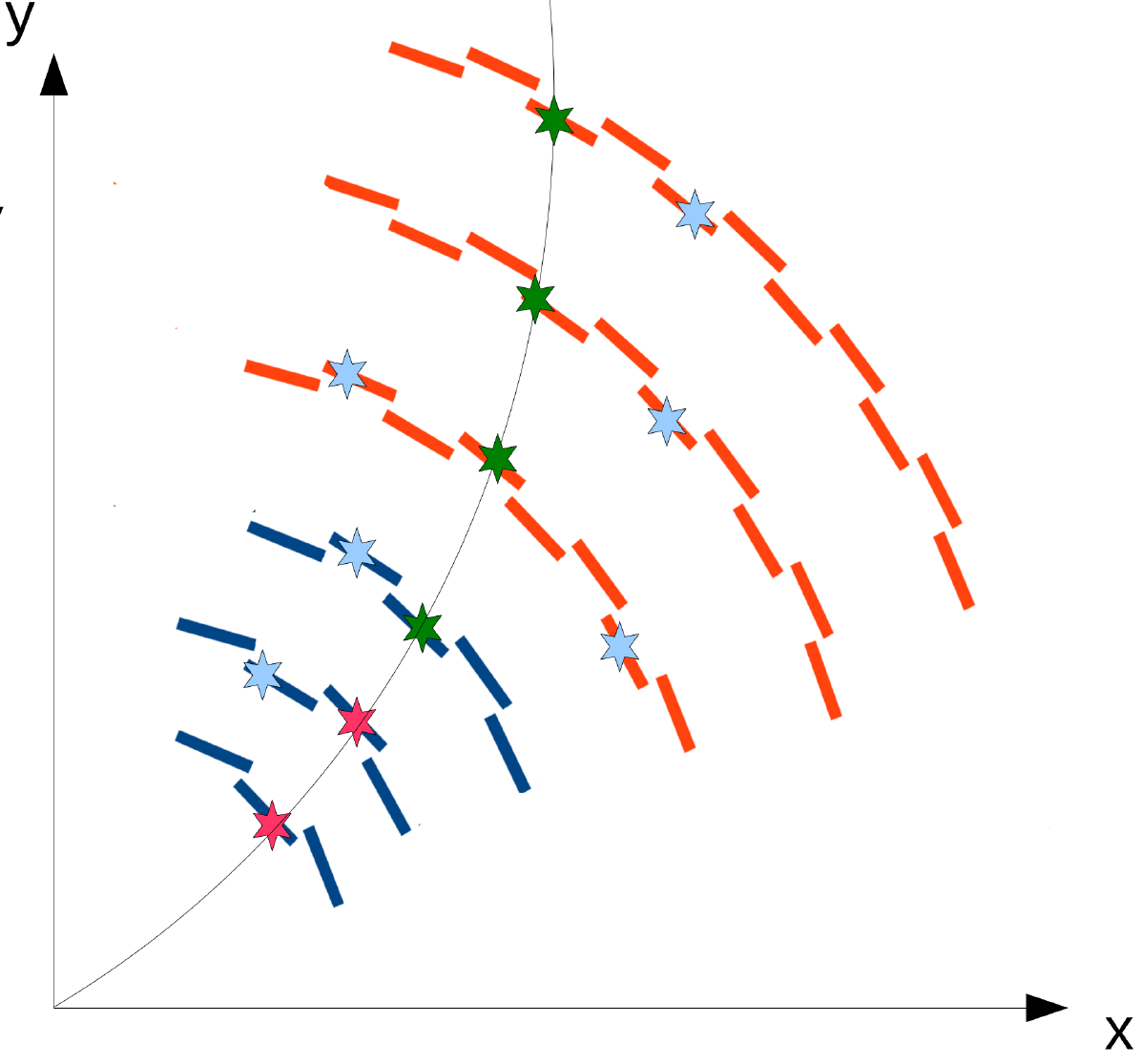}
\caption{\label{fig:trackletconcept} Illustration of the various steps of the tracklet algorithm. In the first step (left) pairs of
  stubs (red stars) are combined to form seeds, or tracklets, for the track
  finding. Combined with the interaction point (0,0) a helical
  trajectory for the particle is formed, assuming a uniform magnetic
  field.  This trajectory is projected (middle) to the other
  layers. Stubs in the other layers that are close to the projection
  (green stars) are selected as matches (right) to the tracklet to form a
  track. Final trajectory parameters are calculated using a linearized
  $\chi^2$ fit. }
\end{center}
\end{figure*}

The tracklet algorithm proceeds in multiple steps, illustrated in Figure~\ref{fig:trackletconcept}. 
The algorithm begins with a seeding step, where tracklets are
formed from pairs of stubs in adjacent layers or disks. An initial estimate of the tracklet
parameters is calculated from the two stubs and using the detector
origin as a constraint in the $r$--$\phi$ plane.  Seeds are rejected
if they are inconsistent with a track with $\pt > \unit[2]{GeV}$ and
$|z_0| < \unit[15]{cm}$. 
The seeding is performed in multiple pairs of layers (or disks) to ensure 
good efficiency for full azimuthal coverage and for redundancy in the system. 
Nominally, seeding between layers 1+2, 3+4, 5+6, between disks 1+2,
3+4, and between layers 1 or 2 and disk 1, are considered. 
Different seeding layers can be used in the event of malfunction of a particular layer or disk. 

The tracklets are projected to other layers and disks to search for
matching stubs. These projections use predetermined search windows,
derived from measured residuals between projected tracklets and stubs
in simulated data. The tracklets are projected both inside-out and
outside-in, i.e., towards and away from the collision point, as needed
for a given seeding combination. If a
matching stub is found, the stub is included in the track candidate
and the difference between the projected tracklet
position and the stub position, ``residual'', is stored. If there
are multiple stubs matched in a given layer or disk, the stub with the
smallest $\phi$ residual is retained for the track fit.

A linearized $\chi^2$ fit is performed for all stubs matched to the trajectory. The track fit implementation uses pre-calculated derivatives and the tracklet-stub residuals from the projection step. The linearized $\chi^2$ fit corrects the initial tracklet parameters to give the final track parameters: inverse radius of curvature ($\rho^{-1}$), azimuthal angle ($\phi_0$), polar angle ($\tan\theta$), $z$ intercept ($z_0$), and optionally the transverse impact parameter ($d_0$).  
The main source of duplicates is due to the same track, with the same hits, being found multiple times using different seeding layers. The initial duplicate removal we studied wanted to identify these duplicates and the use of the $\chi^2$ from the fit was not initially considered. However, with further studies we have, in tracklet 2.0, now also considered the use of the $\chi^2$ in the duplicate removal.
%
%

When implementing the algorithm on an FPGA, we work with fixed-precision math and low-order Taylor expansions of trigonometric functions. The number of bits kept are adjusted to ensure adequate precision. 
The loss of precision from using fixed-point calculations is negligible. 


\subsection{Parallelization}
\label{sec:parallelization}

The algorithm is parallelized in the following manner. First, the detector is split along azimuth into sections,
called ``sectors''. The number of sectors is a tunable parameter and is chosen based on an optimization of the cost of FPGAs, the required per-sector input bandwidth, constraints due to the cabling interface of the on-detector electronics, the choice of time-multiplexing factor and the overall algorithm latency, as discussed in the following paragraphs. The nominal choice for the number of sectors is nine, while other values were also considered. For a given bunch crossing, each sector is assigned a dedicated hardware called the sector processor, and tracklet formation and track reconstruction are done in parallel in all of the sector processors. A small amount of data is duplicated near the boundaries of sectors to allow track formation to take place entirely within one sector processor and to avoid gaps in detector coverage. The data duplication scheme will be described in Section~\ref{cabling}. 

The upper limit on the number of sectors is chosen such that a track with largest acceptable curvature ($\pt=\unit[2]{GeV}$) is contained in at most two sectors. This corresponds to 28 sectors, and was the configuration used for the tracklet 1.0 hardware demonstrator. A
minimal overlap for stubs in the even layers and disks were used, such that we 
can form the tracklets within a sector and maintain full seeding
coverage. However, for matching stubs to projections in tracklet 1.0
we used a nearest neighbor sector communication for the 
projections that extended outside the sector. In addition to the
projections that were sent to the nearest neighbor, matches
that were found also needed to be sent back. This means that we
had two steps in the algorithm where communication with neighboring
sectors were needed. For the tracklet 2.0 implementation, the sector definition was 
revised to instead use 9 $\phi$-sectors. These sectors are 
defined such that they include all stubs required to reconstruct 
tracks that pass through the sector at $r_{\rm crit}=\unit[55]{\mathrm{cm}}$. This 
avoids the need for the nearest neighbor communication and, as a consequence, 
saves approximately \unit[1]{\mus} of latency. 

The system of sector processors is replicated $n$ times using a round-robin time multiplexing approach. Each system is entirely independent, and therefore, since new data are generated every \unit[25]{ns}, each independent time multiplexed unit has to process a new event every $n\times \unit[25]{ns}$.  As with the number of sectors, the choice of time multiplexing factor $n$ is driven by a balance of cost, efficiency, and needed processing power.
For the current system, $n=18$ is considered to balance these three factors; that is, each sector processor receives new data every \unit[450]{ns}.  For the 28 sector configuration, $n=6$ was chosen, leading to each sector receiving data every \unit[150]{ns}.

By construction, the system operates with a fixed latency.  Each processing step proceeds for a fixed amount of time. If we have too many objects, some will not be processed, leading to truncation of processing and an algorithmic inefficiency.

The algorithm is further parallelized within sectors.  In the serial algorithm, there are several places where loops over stubs or double loops over pairs of stubs are required. In a naive implementation, the time to process these parts of the algorithm scales as $N$ or $N^2$, where $N$ is the number of stubs in the sector, if considering all possible combinations. The number of combinations, or combinatorics, is a challenge to the algorithm.
The combinatorics in forming tracklets and matching tracklet projections to stubs is efficiently reduced by dividing sectors into smaller units in $z$ and $\phi$ to allow additional parallel processing. These smaller units are referred to as ``Virtual Modules'' (VMs). Only a small fraction of VM pairs can form a valid tracklet -- the majority would be inconsistent with a track originating at the point of collision and with high enough transverse momentum.  Data are distributed into those VMs satisfying these requirements in an early stage of the algorithm.  This subdivision efficiently reduces the number of combinations that need to be considered by the algorithm from the start. Additionally, each VM is processed in parallel. At the next stage of the algorithm, the amount of parallelism is reduced when the accepted VM pairs' (the tracklets) initial track parameters are calculated.

Two different configurations of VMs are considered. 
In the first, used in tracklet 1.0 for  the demonstrator, a sector is divided into $8z\times3\phi$ ($8z\times4\phi)$ VMs for the odd (even) layers, resulting in 24 (32) VMs per layer.  In forming the tracklet seeds, only 96 of the 768 ($=24\times32$) possible combinations are consistent with a possible tracklet that has the appropriate $z_0$ and maximum curvature (minimum momentum); the others do not need to be considered. It is by this method of data binning that the VMs eliminate the need to expend cycles considering combinations that are known a  priori not to lead to a viable tracklet candidate.

In the second configuration (long VMs) used in tracklet 2.0, a finer $\phi$ segmentation is used. The sectors are here subdivided into $24\phi$ ($32\phi$) VMs for the odd (even) layers. These VMs cover the full length of the sector but, internally to the VM, the data are collected into eight $z$ bins. With this configuration, 120 of the same 768 combinations lead to viable tracklet seeds. The advantage of this configuration is that it leads to more even resource usage (better load balancing) and the binning in $z$ allows a significant reduction in the combinatorics that has to be tried when forming tracklets and matching projections to stub, as will be demonstrated in Section~\ref{sec:demonstrator}.  In the firmware implementation, the difference between the two versions of the algorithm is restricted to a few places in the design.


%% file: firmware.tex
\section{Firmware implementation}
\label{sec:firmware_impl}

\begin{figure}
\begin{center}
\includegraphics[width=0.99\textwidth]{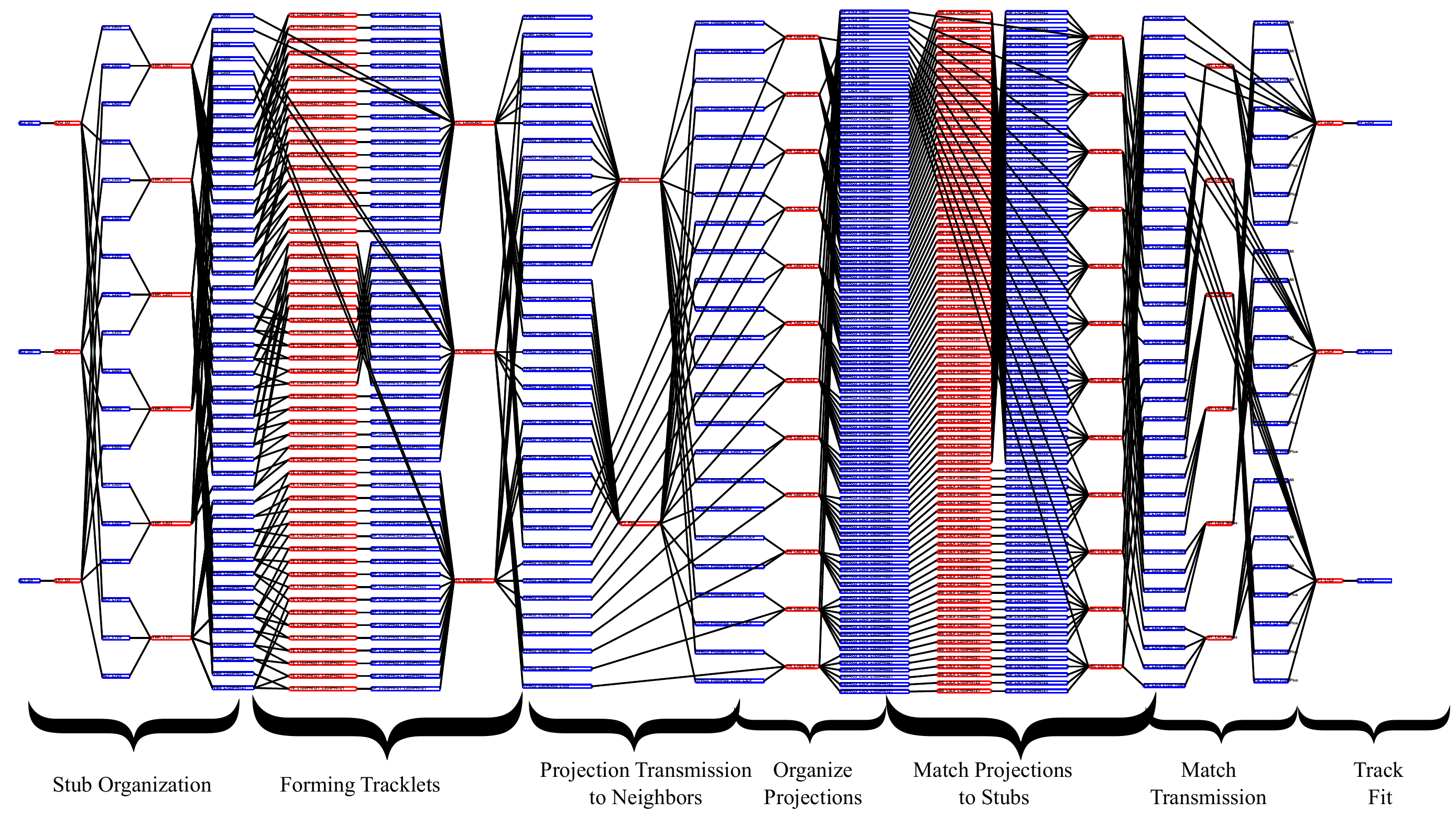}
\caption{\label{fig:trackletproject} An illustrative high-level routing diagram for the tracklet project. As described in the text, the project consists
  of nine processing steps and two transmission steps indicated in
  red, and memories, shown in blue, in which the data are stored
  between the processing steps. The black lines show which processing
  block reads (and writes) from which memory. The input stubs arrive
  from the left in the picture and the first two processing modules
  sort the stubs into the correct VMs based on the
  physical location of the stubs. The next two stages involve the
  tracklet finding and projection calculation. Next the projections
  are routed to the correct VMs and then the projections
  are matched to stubs and the matches are then used to perform the
  track fit.  The removal of duplicates is performed after the track fit (not shown for reasons of clarity).  }
\end{center}
\end{figure}

The tracklet algorithm is implemented in the \textsc{Verilog} Hardware Description Language (HDL)~\cite{verilog} as nine processing steps and two transmission steps~\cite{jorge}. These
processing steps are illustrated in Figure~\ref{fig:trackletproject}. The red boxes are processing modules
and the data are stored in memories (blue boxes) between the different processing steps. The black lines indicate which processing modules read and write data from which memory. 
The implementation of the algorithm in the FPGA takes place in the following  steps.
\begin{itemize}
\item {\it Stub organization:} (1) Sort the input stubs by their corresponding layer ({\bf LayerRouter}), and (2) into smaller units in $z$ and $\phi$, the VMs ({\bf VMRouter}).
\item {\it Tracklet formation:} (3) Select candidate stub pairs for the formation of tracklets ({\bf TrackletEngine}), and (4) calculate the tracklet parameters and projections to other layers ({\bf TrackletCalculator} module).
\item {\it Projections:} (5) Transmission of projections pointing to neighboring sectors ({\bf ProjectionTranceiver}). (6) Route the projections based on smaller units (VMs) in $z$ and $\phi$ ({\bf ProjectionRouter}).
\item {\it Stub matching:} (7) Match projected tracklets to stubs ({\bf MatchEngine}), and (8) calculate the difference in position between the stubs and projected tracklet ({\bf MatchCalculator}). (9) Transmission of matches between sectors ({\bf MatchTransceiver}).
\item {\it Track fit:} (10) Perform track fit; update the initial tracklet parameter estimate ({\bf TrackFit}).
\item \emph{Duplicate Removal:} (11) Remove tracks found multiple times (\textbf{PurgeDuplicate}).
\end{itemize}
Both the tracklet 1.0 and tracklet 2.0 projects follow the same structure; they differ in the internals of some of the modules and how the modules are inter-connected.

Each of the steps outlined above corresponds to HDL modules (named in
bold). These modules are hand-optimized. They can be customized
with \textsc{Verilog} parameter statements on instantiation to account
for differences between use cases. For example, in the second step of
stub organization, six router modules are needed to process the stubs
in each layer. The bit assignment in the data differs between the
inner and outer three layers of the barrel. On instantiation, a
parameter is used to select the appropriate version.
The project illustrated in Figure~\ref{fig:trackletproject}
corresponds to \sfrac{1}{4} of the barrel for one sector for the tracklet 1.0 configuration. A complete project would contain approximately eight times as many instantiations
of the same set of modules.  The wiring between modules is specified
in a master project configuration file. This configuration file is
processed with \textsc{python} scripts to generate the top-level
\textsc{Verilog}, which is then synthesized using Xilinx Vivado
2016.1. These \textsc{python} scripts also generate the module
connection diagram shown in Figure~\ref{fig:trackletproject} and
drive a bit-level \textsc{C++} emulation of the system.

All processing modules follow a similar format where the input is read from memories filled by the previous step and the output is written to another set of memories.  All processing modules across all sector processors use a single common clock (currently, \unit[240]{MHz}). As soon as a new event arrives, the next step in the chain will start processing the previous event. This implies that at any given time, several events are in the processing pipeline depending on the number of processing steps. Though the project as implemented in the demonstrator used the same clock for all processing steps, the fact that we use memories as buffers between the different steps allows the use of different clock speeds for different processing modules.

An event identifier  propagates with the data and  is
used by the processing steps to access the appropriate data.
We use the event identifier in the top bits of the memory
address. This assumes a fixed maximum number of entries per event in
the memory buffer. The fixed latency design implies that the maximum
number of entries that can be processed is known and as such the
limitation due to the fixed number can be understood and tuned. Most
of the data from a processing step is only used in the next step and thus
we can make very shallow buffers that will hold only two events at the
same time (writing one and reading the other). These small buffers are
implemented as distributed RAM in order to minimize the Block RAM (BRAM) resource usage in the FPGA. On the other hand, some data need
to be stored for up to eight events since they will only be used later
in the chain. These data are stored in BRAMs, but we try to minimize the
usage of this resource as we have observed correlation of routing
difficulties with the number of BRAMs used.

Since the calculations needed for routing the data are simple and
using Lookup Tables (LUTs) is quick, most of the processing modules
take only a few clock cycles to complete. We do not send the data to
the next step immediately, but buffer it in memories until the
allocated time is finished for the processing step. At this time, the
module corresponding to the next step in the processing will start
reading the data for the previous event and new data will be written
for the current event. We use the true dual-port memories available in
the Xilinx FPGAs for our buffers such that we can write the data from
one event while simultaneously reading from the previous one. These
dual-port memories also allow different modules to exist in separate
clock domains.

In addition to the nine processing modules, we also implement two
steps of neighbor communication using optical high-speed serial links. 
The bending of charged particles in the magnetic field can cause trajectories to curl into neighboring sectors.  In this instance, the projected position of the track is sent across fiber links to the neighbor sector processor to look for matching stubs. Simultaneously as each sector processor is sending data to its left and right neighbors, it is also receiving from them as well for the same purpose. This system configuration was chosen to reduce the amount of data duplication globally at the cost of some increase in latency. As discussed in Section~\ref{sec:algo}, in the tracklet 2.0 implementation nearest neighbor communication is not needed.

\subsection{Module examples}

\begin{figure}
\begin{center}
\includegraphics[width=0.45\textwidth]{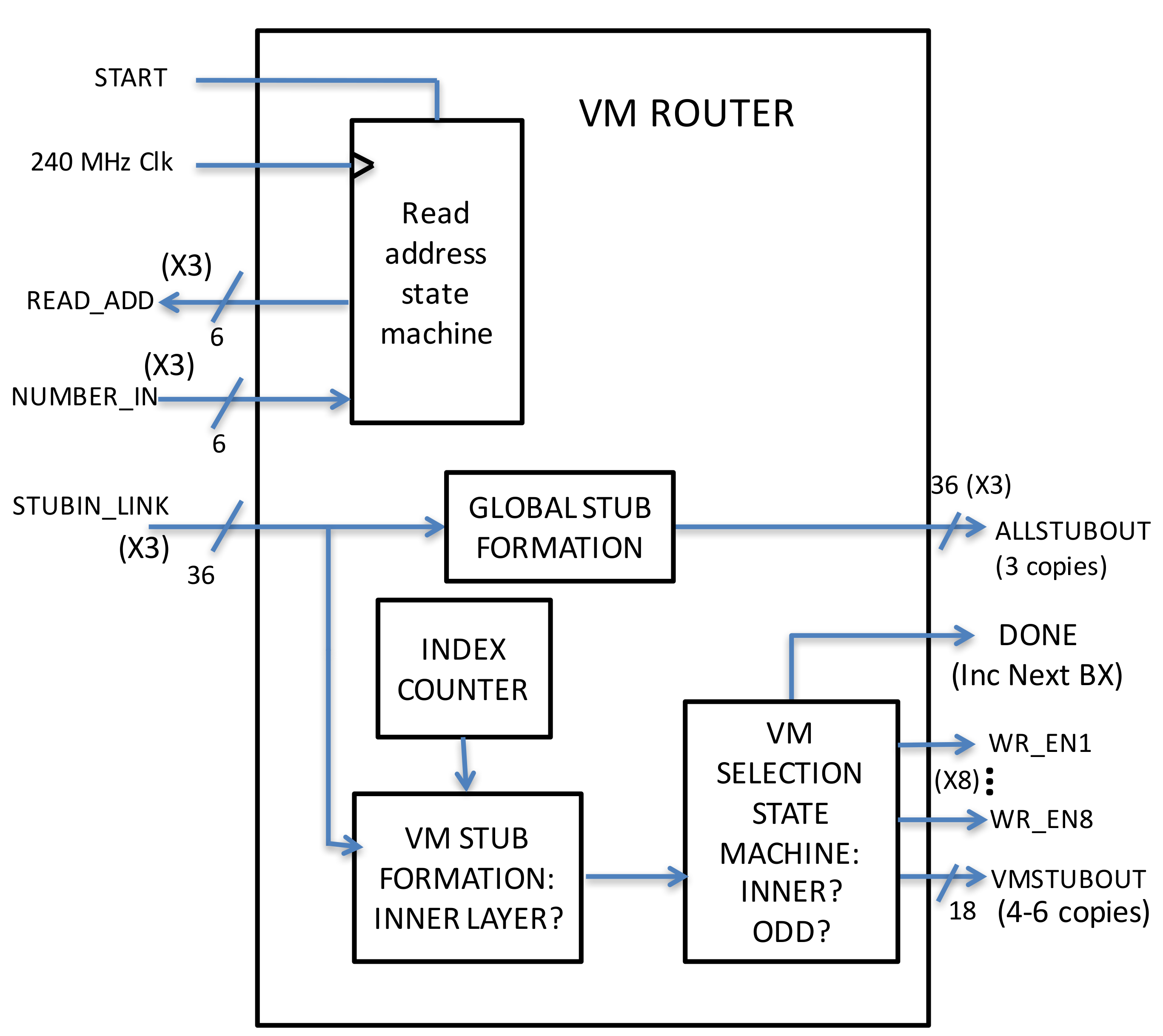}
\caption{\label{fig:vmrouterdiagram}
Schematic illustrating the connections of the VMRouter processing 
module. The module reads input stubs from the
input memories and routes them to the correct VMs based
on the stubs' coordinates ($z$ and $\phi$). A stub per clock cycle is
read from the input stub memories and based on the most significant
bits of the $z$ and $\phi$ coordinates the stub information is written
to the relevant VM memory.
}
\end{center}
\end{figure}

To illustrate the method in more detail, we present  the
functionality of two processing modules. 
Figure~\ref{fig:vmrouterdiagram} shows schematically how the VMRouter works. 
The module receives 
a start signal every \unit[150]{ns} for every new event. This VMRouter module
reads stubs from three input layer memories. 
All stubs are written to the ``AllStubs'' memory in
the full 36-bit format.  In addition, based on their coordinates
($\phi$ and $z$), the stubs are routed to a specific output memory (VM
stub memory) corresponding to a specific small area of the
detector. Here, only coarse position information is retained and a six
bit index into the AllStubs memory is saved such that we can later
retrieve the precise stub position. The process loops over the input
stubs and writes them out to different memories based on their position
information.

\begin{figure}
  \begin{center}
    \includegraphics[width=0.5\textwidth]{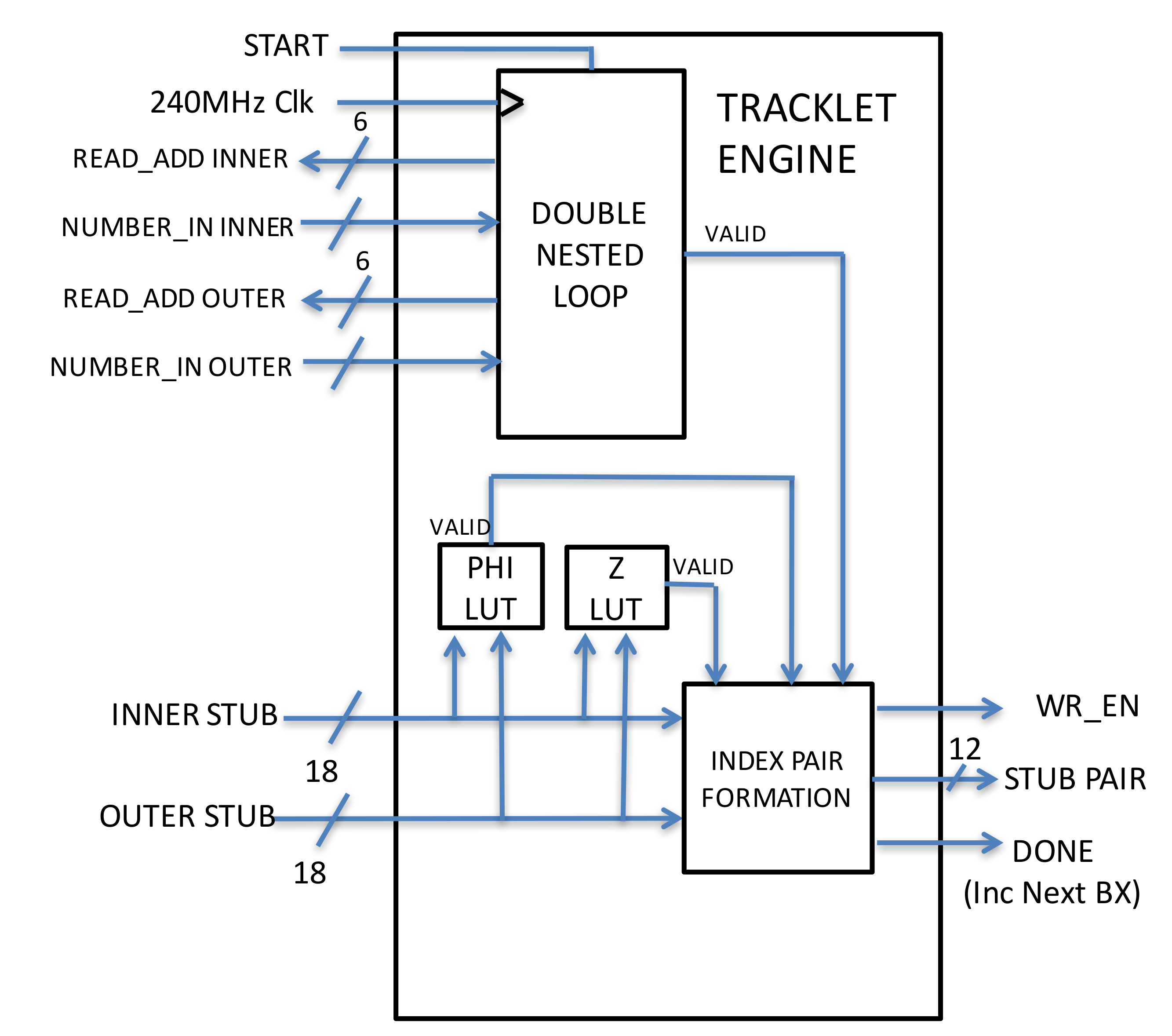}
    \caption{\label{fig:tediagram} Schematic illustrating the
      connections of the TrackletEngine processing module. The module
      reads stubs from two VM memories. Two lookup tables
      are used to check consistency with the momentum and $z$
      vertex. If the pair of stubs pass the selection, a stub-pair
      tracklet candidate is written out.  }
  \end{center}
\end{figure}

A more complex example is the TrackletEngine processing module illustrated in Figure~\ref{fig:tediagram}. This module forms pairs of
stubs as seed candidates. As such, this module
reads input stubs from two VM stub memories filled by the VMRouter
module described previously, but since we are interested in forming
pairs of stubs, this module implements a double nested loop over all
pairs. For each pair the coarse position information is used
in two LUTs to check that the seed candidate is consistent with a
trajectory with the \pt\ and  $z_0$ requirements described above. If the
stub pair passes this check, the indices of the stubs in the AllStub
memories are saved in the output memory of candidate stub pairs.
These indices are used in the next step, the TrackletCalculator, to
retrieve the stubs and calculate the precise trajectory.

\begin{figure}
\begin{center}
\includegraphics[width=0.65\textwidth]{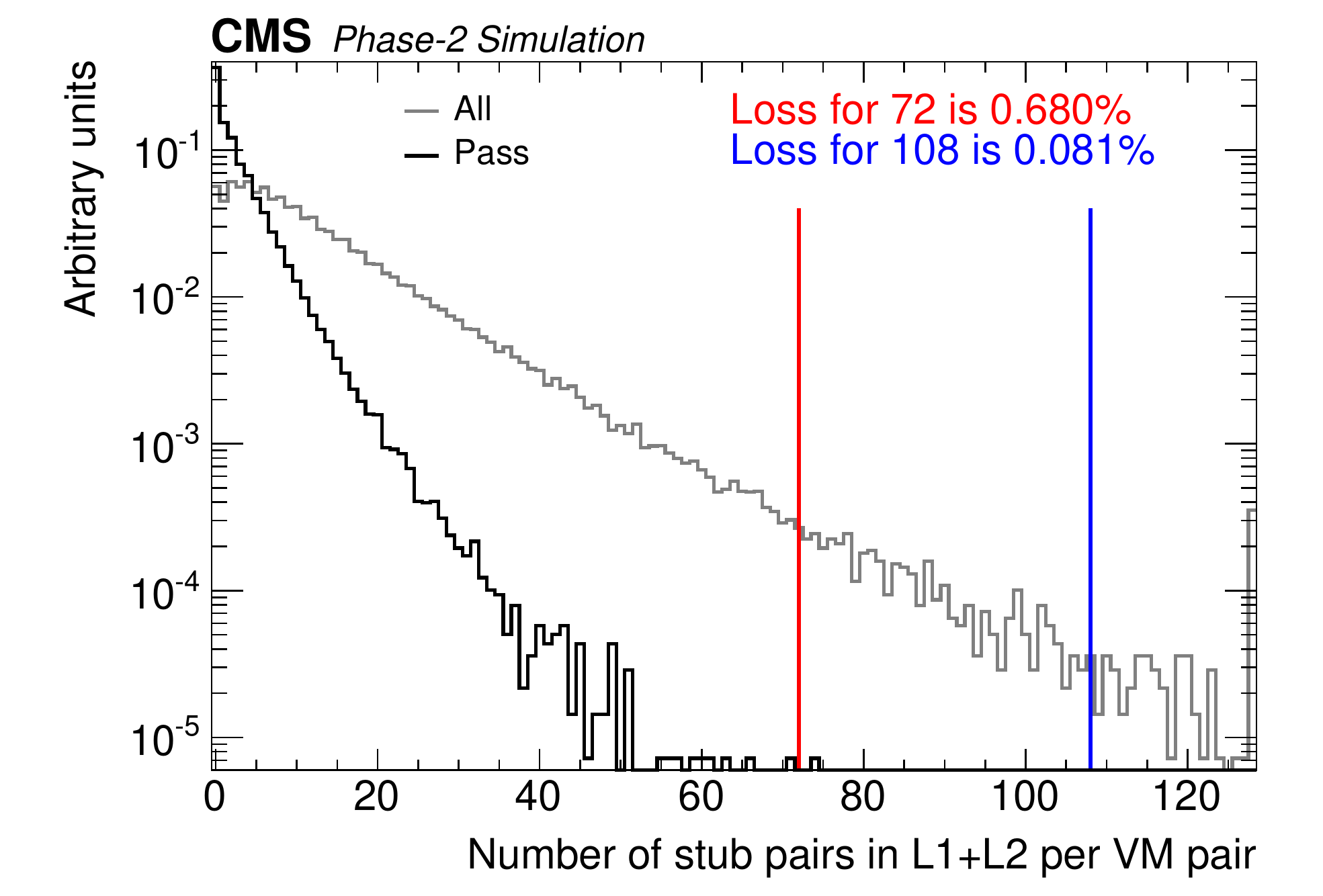}
\caption{\label{fig:tecombinatorics} Simulation of the distribution of
  the number of stub pairs that TrackletEngines seeding in the two innermost barrel layers (L1+L2) have
  to process for $\ttbar$ events with an average pileup of 200. The grey curve shows the number of stub pairs that the
  module has to consider, while the black curve the number of stub pairs that
  pass. The blue line corresponds to 108 processing steps per bunch crossing and
  the red line to 72 processing steps. With a cut-off at 108
  processing steps, we drop less than 0.1\% of the stub pairs. 
}
\end{center}
\end{figure}

Figure~\ref{fig:tecombinatorics} shows the distributions of the number
of stub pairs that each TrackletEngine has to process. Since each step
operates with a fixed latency, we have a maximum number of stub pairs
that can be processed per event. With \unit[450]{ns} per event and a clock
speed of \unit[240]{MHz}, a maximum of 108 input stub pairs can be
considered. As can be seen in the figure, there are cases where there
are more than 108 input stubs; the 109th stub-pair and later will not be
processed and could lead to an inefficiency of the tracking
algorithm. However, due to the built-in redundancy of seeding in multiple
layers, the ultimate effect of this truncation on the final efficiency
is observed to be small, as discussed in Section~\ref{sec:performance}.

%% file: demonstrator_future.tex

\section{Demonstrator system}
\label{sec:demonstrator}
To explore the feasibility of the tracklet system, a demonstrator was completed in 2016. The goal of the test was to implement an end-to-end working system, using simulated data, to validate the design methodology and system modeling. The hardware used for the demonstrator system were \utca boards with a Xilinx Virtex-7 (XC7VX690T-2) FPGA~\cite{virtex7} and a Xilinx Zynq-7000 SoC for configuration and outside communication. These so-called Calorimeter Trigger Processor (CTP7) boards~\cite{ctp7} were developed for the current CMS trigger~\cite{cms_trigger,L1TDR}.  Implementing a full sector in one FPGA on a processing board was out of reach for the older Virtex-7 class FPGAs. For this reason, the demonstrator system focused on the implementation of a half-sector. The simulated data were derived from a \textsc{Geant}-based simulation of the CMS detector~\cite{geant4}. Data sets used included single particle (electron, muon, and pion) events, as well as fully simulated top quark-antiquark (\ttbar) events. To accurately simulate HL-LHC conditions, up to 200 extra \pp collisions were included in addition to the event under study. 

\begin{figure}
  \begin{center}
    \includegraphics[width=.5\textwidth]{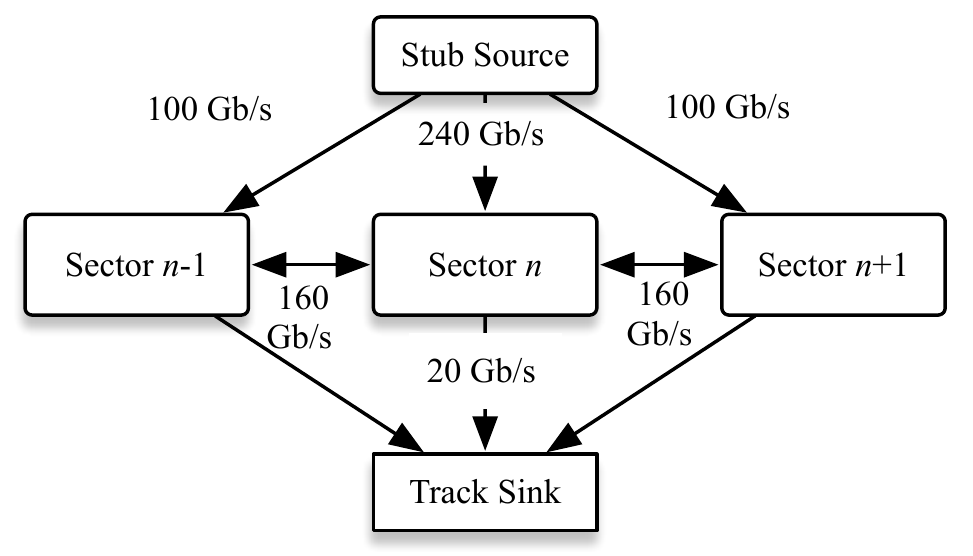}
    \caption{\label{fig:demoschematic} Schematic of the demonstrator
      system.  Three \utca blades implemented three sectors and a
      fourth blade served as the source and sink of data. The central
      sector processor was the actual system under test.}
  \end{center}
\end{figure}

\begin{figure}
\begin{center}
	\includegraphics[width=0.4\textwidth,angle=180]{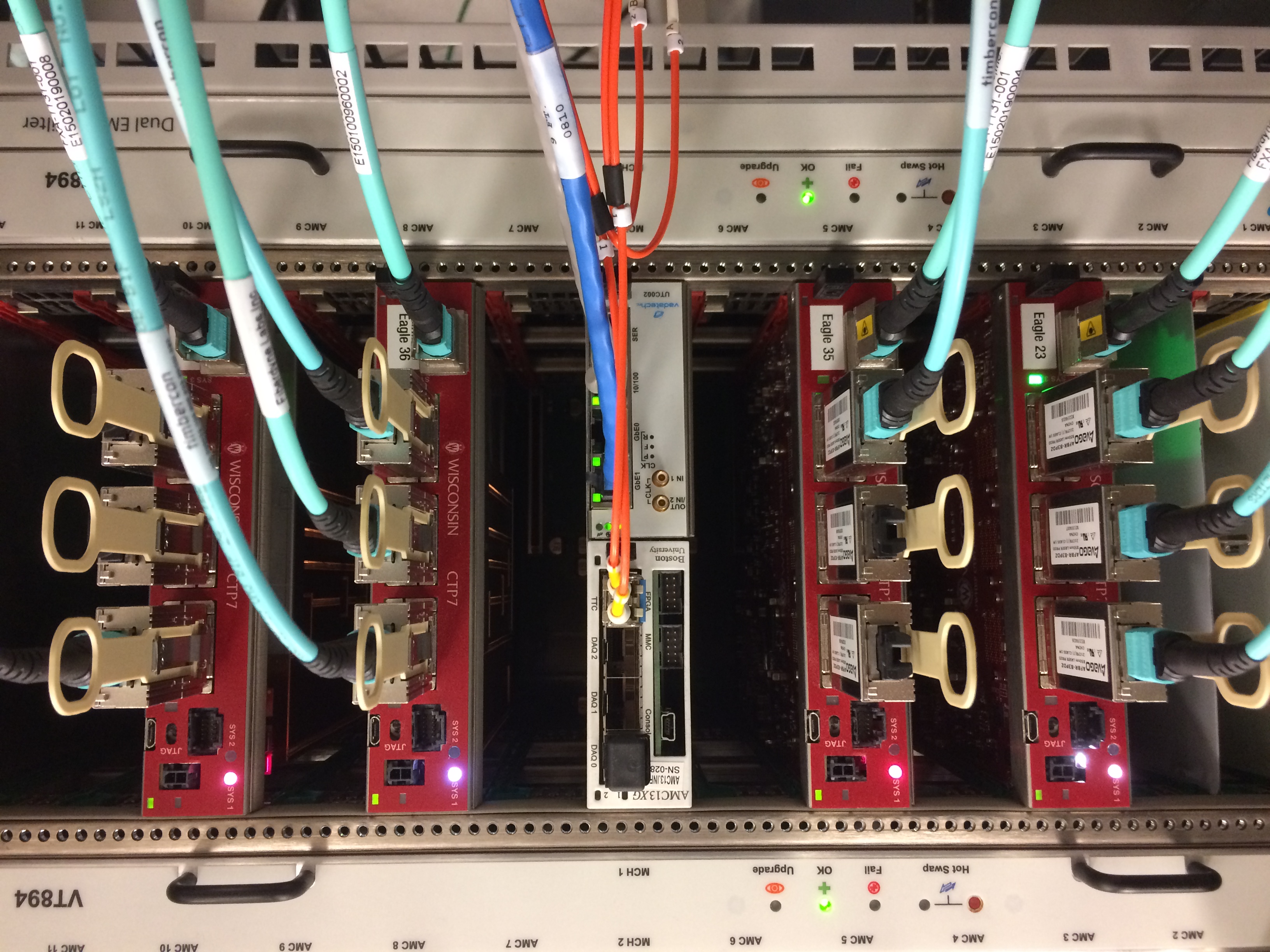}
	\caption{\label{fig:demosystem} The  demonstrator test system. The system was based on the CTP7 \utca blade, used in the current CMS trigger. The system consisted of four CTP7s. Each  CTP7 has 63 input and 48 output \unit[10]{Gb/s} optical links.  }
\end{center}
\end{figure}

The demonstrator system consisted of three $\phi$ sectors and one time-multiplexing slice. A total of four \utca  processing blades were used, one for the central $\phi$ sector, two for its nearest neighbor sectors, and one blade that acted as a data source (providing input stubs) and a data sink (accepting the final output tracks). The system configuration is shown in Figure~\ref{fig:demoschematic}. The demonstrator was fed with simulated data derived from the  simulation of the CMS detector mentioned above. An Advanced Mezzanine Card (AMC13)~\cite{amc13} provided the central clock distribution. The inter-board communication used 8b/10b encoding with \unit[10]{Gb/s} link speed. The demonstrator system is shown in Figure~\ref{fig:demosystem}. The demonstrator system assumed 28 sector processors and a time multiplexing factor of six, leading to new data arriving every \unit[150]{ns}.

Two complete implementations of the firmware project -- one for half the barrel ($+z$) and one for a quarter of the barrel plus the forward ($+z$) endcap -- were used to demonstrate the feasibility of this approach for the full $\eta$ range of the detector.  These two projects covered two regions of the detector:
\begin{enumerate}
\item The barrel-only region. These tracks only traverse the
  cylindrical part of the detectors.
\item The hybrid region. Tracks that traverse both a part of the barrel region as well as the disk region  of the detector.
\end{enumerate}
By dividing the projects into these two regions, we exercised all  regions of the detector while still working within the constraints of the Virtex-7 FPGAs. 

\begin{figure}
\begin{center}
\includegraphics[width=10cm]{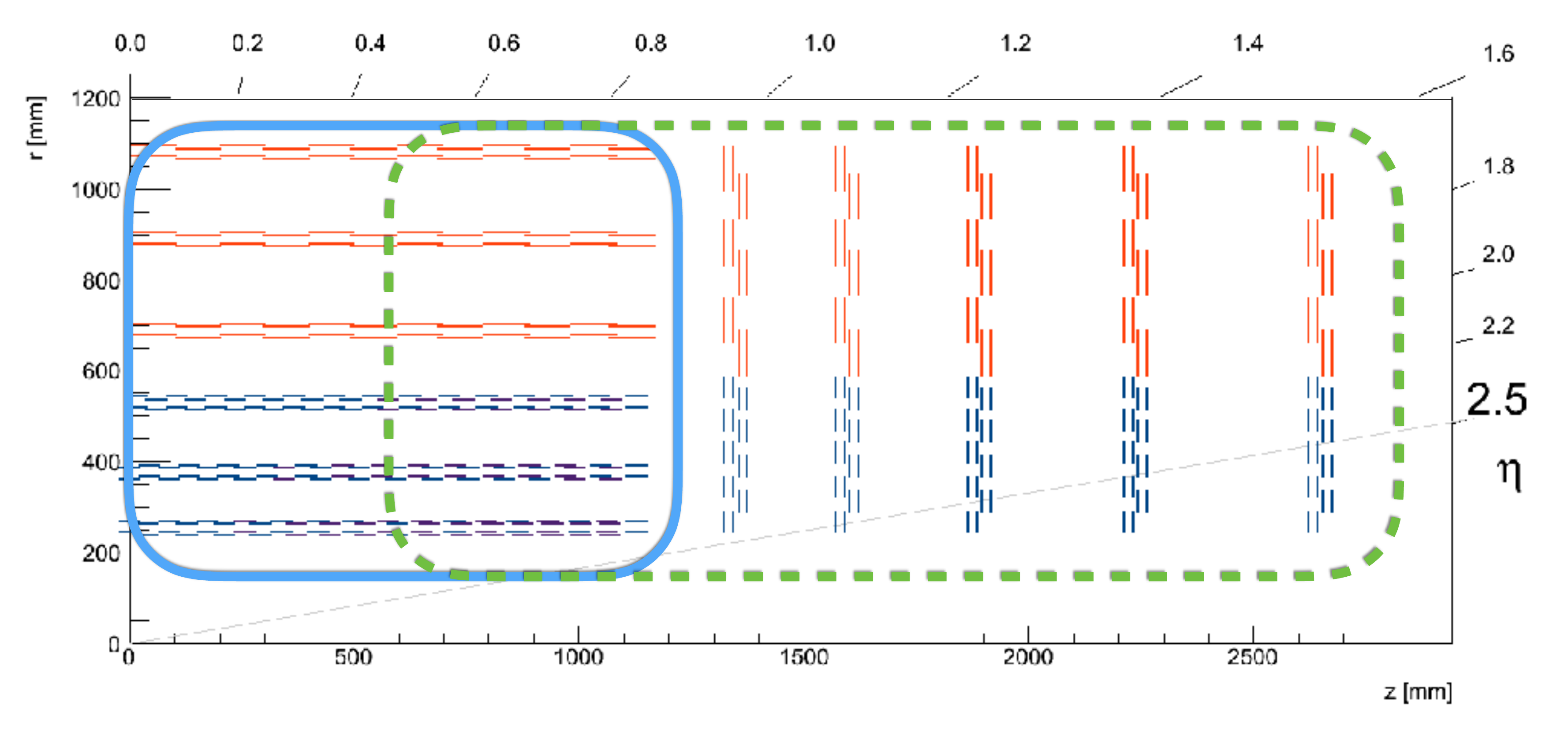}
\caption{Sketch of the $+z$ portion of the detector in the $r$-$z$ plane. A box in blue shows the detector region covered by the half-barrel implementation. A box in green shows the detector region covered by the other project that spans a quarter of the barrel, the transition region between the barrel and endcaps, and the endcaps.}
\label{fig:proj_sketch}
\end{center}
\end{figure}

A sketch of the tracker regions covered by each of the implementations is shown in Figure~\ref{fig:proj_sketch}. As discussed in Section~\ref{sec:firmware_impl}, data formats and calculations are slightly different in the two regions of the detector. The differences were absorbed using parameter statements in the \textsc{Verilog} code and determined at module instantiation time.

\subsection{Validation and testing}

\begin{figure}
 \begin{center}
  \includegraphics[width=0.8\linewidth]{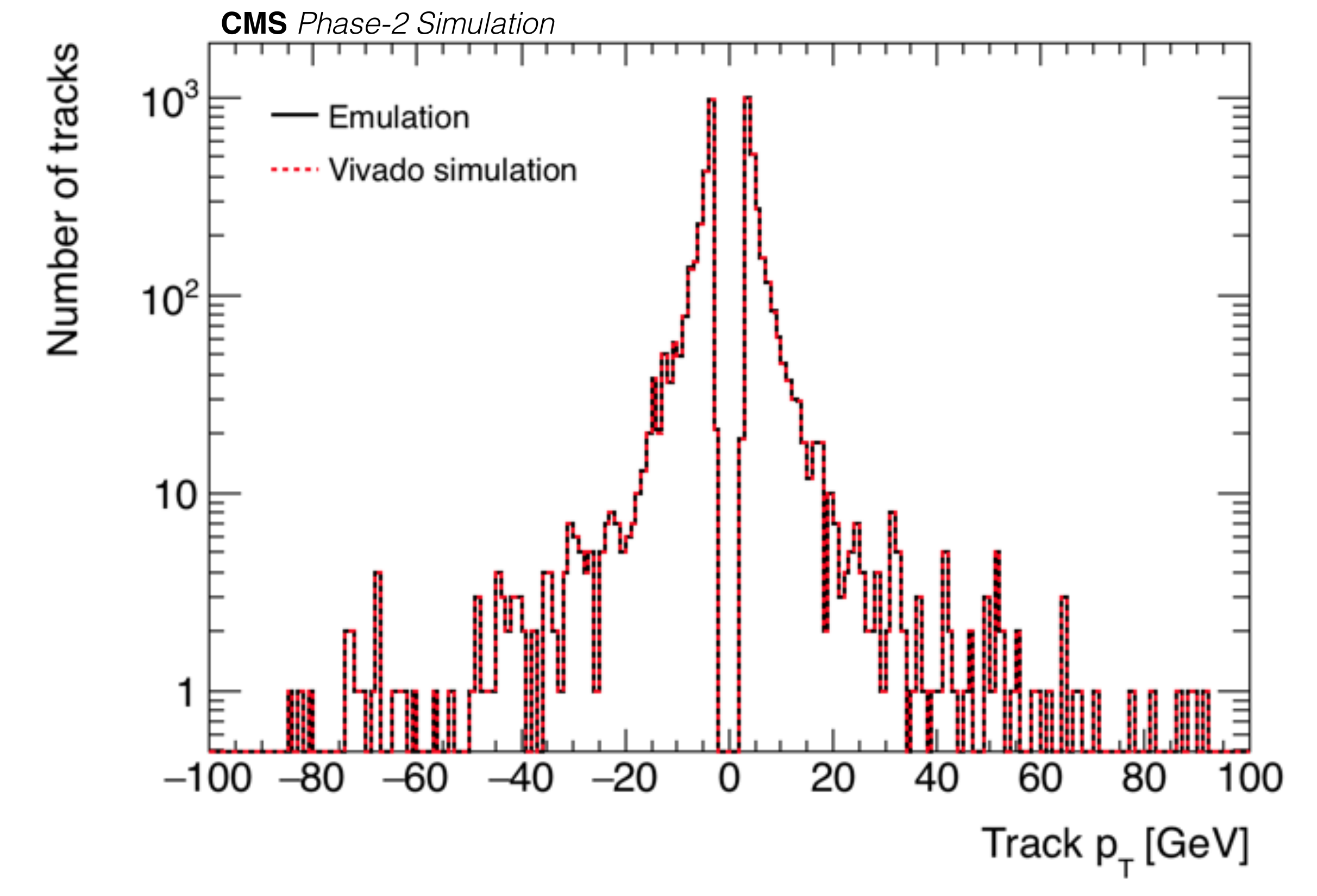}
  \caption{Comparison of the final track $\pt$ between \textsc{C++} emulation and HDL simulation, for simulated \ttbar events with 200 pileup events. Better than 99\% agreement is observed.}
  \label{fig:sim_demo_comparison}
  \end{center}
\end{figure}

Events were processed through the demonstrator as illustrated in Figure~\ref{fig:demoschematic}. First, input stubs obtained from simulations were written to a piece of hardware that emulates the data delivery in the final system. On a GO signal, stubs were sent to the three sector processor boards. A new event was sent to each sector board every \unit[150]{ns}. The events were processed and projections and matches were sent to and received from neighboring boards. The final output tracks were received by the track sink board. Systematic studies were performed to compare the integer-based emulation of the tracklet algorithm with a HDL simulation of the FPGA using Xilinx Vivado, as well as with the output tracks from the demonstrator system. Full agreement was observed in processing single-track events between the emulation, FPGA simulation, and board output. Better than 99.9\% agreement was observed with many-track events with high pileup (Figure~\ref{fig:sim_demo_comparison}). The demonstrator had a 28-fold azimuthal symmetry (i.e., 28 sectors), so we tested the full $+z$ range by using different input data, corresponding to the different sectors, without any modifications of the demonstrator itself.  

\subsection{System latency}

\label{sec:latency}
For both the demonstrator system as well as the current configuration, each processing step of the tracklet algorithm takes a fixed number of clock cycles to process its input data. The processing modules' latency from receiving upstream data to producing the first result varies between $1$--$50$ cycles depending on the module. Each module then continues to handle the data of the same event and writes to the memories for \unit[150]{ns} (for a time multiplexing factor of six) before switching to the next event.
For some of the steps where data transmission between the neighboring sectors was necessary, latency due to inter-board links was also included.  The measured transmission latency was \unit[316.7]{ns} (76 clock cycles), which included serialization and de-serialization via the Xilinx GTH transceivers, data propagation in \unit[15]{m} long optical fibers, channel bonding, and time needed to prepare and pass data from processing modules to the transceivers.
The total latency of the algorithm was therefore the sum of the
processing module latencies and processing time, as well as
inter-board data transmission latency, of all the processing
steps. The latency of the hardware demonstrator also included the data
transmission latency for receiving stubs from and sending final tracks
back to the data source/sink blade.  A summary of the estimated
latency is shown in Table~\ref{tab:latency}.
\begin{table}
  \caption{Demonstrator latency model using the tracklet 1.0 configuration. For each  step,
    the processing time and latency is given. For steps involving data
    transfers, the link latency is given.  The model and measured 
    latency agree  within 0.4\% (three clock cycles).}
  \label{tab:latency}
  \centering 
  \begin{tabular}{lrrrrr}
    \hline
    Step & Proc.  & Step         & Step  & Link  & Step  \\
         & time        & latency      & latency & delay  & total \\ 
         & (ns)        & (clock cycles)& (ns)  &  (ns) & (ns) \\ \hline 
    Input link              &    0.0 &    1  &   4.2   &  316.7  &  320.8   \\  
    LayerRouter            &  150.0 &    1  &    4.2   &   -  &  154.2   \\  
    VMRouter               &  150.0 &    4  &   16.7   &   -  &  166.7        \\  
    TrackletEngine         &  150.0 &    5  &   20.8   &   -  &  170.8        \\  
    TrackletCalculator    &  150.0 &   43  &  179.2   &   -  &  329.2        \\  
    ProjectionTransceiver  &  150.0 &   13  &  54.2   & 316.7  &  520.8      \\  
    ProjectionRouter       &  150.0 &    5  &   20.8   &  -  &  170.8        \\  
    MatchEngine            &  150.0 &    6  &   25.0   &   -  &  175.0        \\  
    MatchCalculator        &  150.0 &   16  &   66.7   &   -  &  216.7        \\  
    MatchTransceiver       &  150.0 &   12  &  50.0   & 316.7  &  516.7      \\  
    TrackFit                    &  150.0 &   26  &  108.3   &   -  &  258.3        \\  
    PurgeDuplicate     &  0.0 &   6  &   25.0   &   -  &  25.0        \\  
    Output link             &    0.0 &    1  &    4.2   & 316.7  &  320.8        \\  
\hline 
    Total                   & 1500.0 &  139  &  579.2   & 1266.7  & 3345.8    \\
    \hline
  \end{tabular}
\end{table}
With a \unit[240]{MHz} clock and
a time-multiplex factor of six, the total estimated latency is
\unit[3345.8]{ns}.
The total latency of the demonstrator was also measured with a
clock counter on the data source/sink blade. We started the counter when
sending out stubs, and record the counter outputs when receiving valid
tracks. The actual measurement was done with \unit[240]{MHz} processing clock.
The measured latency was \unit[3333]{ns}, which agrees within three clock
cycles (0.4\%) with the model.\footnote{%
  We did not try to understand the source of the final missing three clock cycles, as an example of the law of diminishing return.
}

For the tracklet 2.0 release, we are implementing the algorithm in
\textsc{C/C++} using the Xilinx Vivado HLS tools~\cite{HLS}. We
believe HLS will produce code that will be easier to maintain and
provide a lower barrier for entry of new people to contribute to the
development of the project.  Tracklet 2.0 provides several
improvements over the algorithm as implemented for the demonstrator.
The algorithmic improvements from using the ``long virtual modules''
reduce the effects of truncation and the new sector definitions remove
the need for nearest neighbor communication.  One significant change
is that the TMUX factor is now 18, which results in \unit[450]{ns} of
processing time. This time increase is offset by the reduction in the
number of processing steps achieved removing the steps that require 
communication with the nearest neighbors. This  allows us to keep
within the \unit[4]{\mus} latency target. Further reductions in
latency are possible by combining processing steps into fewer modules.
The estimated latency with this new configuration is shown in
Table~\ref{tab:latency_hg}.

 \begin{table}[tbp]
  \caption{The estimated latency without sector-to-sector communication 
  and with a time-multiplex factor of 18. The same clock speed (\unit[240]{MHz}) as in the demonstrator is assumed. We assign conservatively \unit[150]{ns} for the link latency, as studies suggest  it could be reduced by \unit[200]{ns} compared to the demonstrator version with  improved link speed and protocol. }
  \label{tab:latency_hg}
  \centering 
  \begin{tabular}{lrrrrr}
    \hline
    Step & Proc.  & Step         & Step  & Link  & Step  \\
         & time        & latency      & latency & delay  & total \\ 
         & (ns)        & (clock cycles)        & (ns)  &  (ns) & (ns) \\ \hline 
    Input link              &    0.0 &    1  &   4.2   &  150.0  &  154.2   \\  
    VMRouter               &  450.0 &    4  &   16.7   &   -  &  466.7        \\  
    TrackletEngine         &  450.0 &    5  &   20.8   &   -  &  470.8        \\  
    TrackletCalculator    &  450.0 &   43  &  179.2   &   -  &  529.2        \\  
    ProjectionRouter       &  450.0 &    5  &   20.8   &  -  &  470.8        \\  
    MatchEngine            &  450.0 &    6  &   25.0   &   -  &  475.0        \\  
    MatchCalculator        &  450.0 &   16  &   66.7   &   -  &  516.7        \\  
    TrackFit                    &  450.0 &   26  &  108.3   &   -  &  558.3        \\  
    PurgeDuplicate     &  0.0 &   6  &   25.0   &   -  &  25.0        \\  
    Output link             &    0.0 &    1  &    4.2   & 150.0  &  154.2        \\  
\hline 
    Total                   & 3150.0 &  113  &  470.8   & 300.0  & 3920.8    \\
    \hline 
  \end{tabular}
\end{table}

The half-sector project included seeding in multiple layer and
disk combinations (L1+L2, L3+L4, D1+D2, and D3+D4). This project
consisted of the following processing modules: 12 LayerRouters, 22
VMRouters, 126 TrackletEngines, 8 TrackletCalculators, 22
ProjectionRouters, 156 MatchEngines, 22 MatchCalculators, 
4 TrackFits, and one PurgeDuplicate.  The resources used in the
demonstrator project are shown in Table~\ref{tab:fpgaresources_2}.  
The resource usage from the \textsc{Verilog} synthesis for
the full sector project is summarized in Table~\ref{tab:fpgaresources}. The
most heavily used resource was BRAMs.  We ran the project at
\unit[240]{MHz}, limited by external constraints. First, we operated
the links at \unit[10]{Gb/s} with a 8b/10b encoding. This meant that
we transferred data packages of \unit[32]{bits} at \unit[250]{MHz}. We
also wanted the clock frequency to be a multiple of the \unit[40]{MHz}
bunch collision rate in CMS. This meant that if we operated at
\unit[240]{MHz} and produced \unit[32]{bit} data packages at this
clock speed the links could transport the data. 
Increases in  clock speed are being considered as a future improvement.  

\begin{table}
  \caption{Summary of post-synthesis resource utilization on the Virtex-7 FPGA for the half-sector project.}
  \label{tab:fpgaresources_2}
  \centering
  \begin{tabular}{lc}
    \hline
Resource type & Estimated utilization (\%) \\
\hline
Lookup tables  & 46 \\
LUT memory  & 29 \\
Flip-flops  & 39 \\
Block RAMs  & 61 \\
Digital signal processing  blocks & 21 \\
Input and outputs  & 8 \\
Gigabit transceivers  & 80 \\
Global clock buffers  & 38 \\
Mixed-mode clock manager  modules & 15 \\
    \hline
  \end{tabular}
\end{table}

\begin{table}[bth]
  \caption{Estimated FPGA resource utilization based on simulation of the
    Tracklet 2.0 full sector project, compared to the resources
    available in the Xilinx Ultrascale+ VU7P device. }
  \centering
  \begin{tabular}{lcccc}
    \hline
    Resource & Unit & VU7P & Tracklet 2.0 & Fraction used (\%) \\
    \hline
    18kbit BRAMs & count & 2810 & 1108 & 39 \\
    DRAM &  Mb & 24 & 4.55 & 18 \\
    288kbit UltraRAMs & count & 625 & 200 & 32 \\
    DSP & count & 4560 & 1696 &  37 \\
    \hline
  \end{tabular}
  \label{tab:fpgaresources}
\end{table}

%


%% file: hardware_impl.tex
\section{Hardware platform and system architecture}
\label{sec:hw_platform}

\begin{figure}
\begin{center}
  \includegraphics[width=0.8\linewidth]{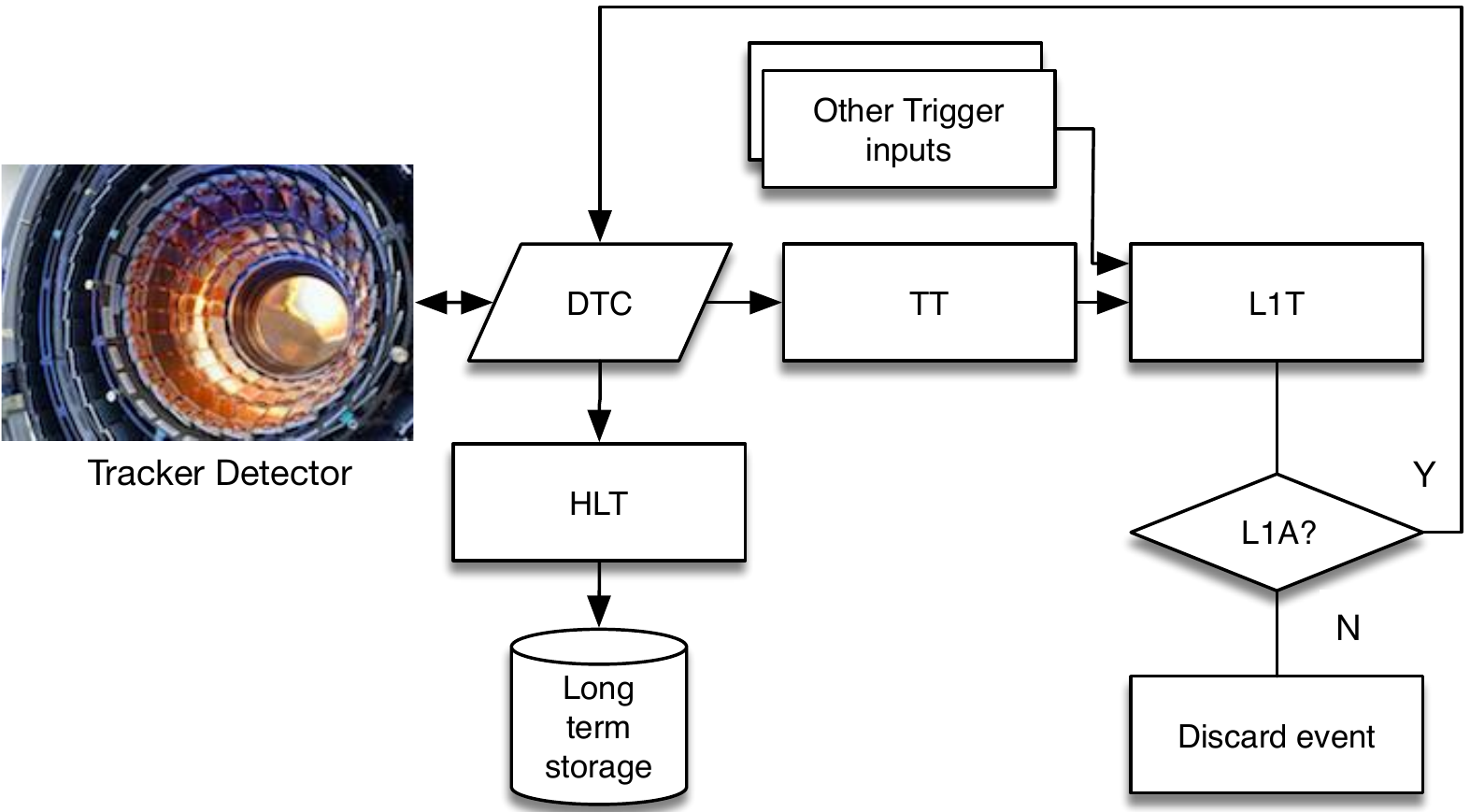}
  \caption{High-level overview of the data flow for the tracker data, showing how the track trigger is integrated into the detector readout and the trigger flow.}
  \label{fig:dataflow}
  \end{center}
\end{figure}

A high-level overview of the tracker data flow is shown in
Figure~\ref{fig:dataflow}. The data from the tracker are sent off the
detector to the Data, Trigger, and Control (DTC) cards~\cite{tracker_tdr}. About 90\% of
the bandwidth is dedicated to a reduced-precision data stream that is
sent to the L1 tracking system, or the ``track trigger'' (labeled TT in
the figure and described in this paper). Data from the track trigger
are sent downstream to a part of the L1 Trigger system (L1T) called the
correlator system, where the information from the tracking detector is
combined with information from other subdetectors to identify
electrons, muons, and other physics quantities\footnote{These
  quantities are used to decide if the event should be dropped or
  stored for further analysis. In total, CMS can store about
  $0.0025\%$ of the collisions for offline study.}. Based on these
data, a L1 Accept (L1A) is issued or not. If the L1A is issued, a
signal is sent to the DTC, which then initiates the readout of the
full-precision data from the detector for that bunch crossing. These data are then sent to secondary software trigger processors (the High-Level Trigger, HLT) for potential storage to disk. If the L1A is not issued, the data are discarded by letting the on-detector data buffers expire.  As can be seen in the diagram, the track trigger receives its data upstream from the DTC cards and sends the results of its processing downstream to the L1 trigger. 

As was mentioned in Section~\ref{sec:parallelization}, the processing of the L1 tracking occurs in sectors that span the entire polar angle and a slice in $\phi$. Each sector is assigned a dedicated hardware unit, the sector processor. The system design centers around these sector processors.  Each sector processor is foreseen to be a single ATCA blade with a Xilinx Virtex Ultrascale+ FPGA. 
The heart of the tracklet approach is this FPGA. It must have adequate DSP resources (about 2000~DSP48E2 equivalent units), I/O (about 50 high-speed serial transceivers running at \unit[28]{Gb/s}), 2800 \unit[18]{kb} BRAMs, approximately \unit[2.5]{Mb} of distributed LUT RAM, and adequate LUT resources.  These requirements are met by the Xilinx Virtex UltraScale+ family.  The input data (stubs) will be received from the upstream DTCs via 24--36 high-speed optical serial links running at \unit[28]{Gb/s}. Output data (reconstructed tracks) will be sent downstream over a single \unit[25]{Gb/s} link. The details of the foreseen cabling for the system is discussed in the section below.

The complete set of sector processors is duplicated a small number of times (time multiplexing of 18 is currently under consideration) and data are distributed to these identical copies in a round-robin scheme. For a system with 9 sectors and a factor of 18 time-multiplexing, the complete system will consist of 162 blades.

The trigger is implemented in custom hardware and sits in a well-shielded cavern away from the detector; as such, radiation and single-event upsets are not a concern.  Initial estimates suggest that the per-slot power requirements will be well below the ATCA specification of \unit[400]{W} per slot.

\subsection{Cabling}
\label{cabling}

The baseline strategy for the project assumes a tilted barrel
geometry~\cite{tracker_tdr}, 9 $\phi$-sectors for track finding and
a time-multiplex factor of 18 for the tracklet 2.0 configuration. 
The DTC layout assumed here
is as follows: a total of 108 DTCs for a half-detector ($+$ and
$-z$); 54 for the PS modules, connected to the inner three layers of the half-barrel and the inner seven rings of the endcap, and 54 for the 2S modules, connected to the outer three layers of the half-barrel and the outer five rings of the endcap.
The DTCs are arranged such that they cover ``nonants'' in $\phi$ (nonants is referring to the currently proposed division of the readout of the outer tracker in a nine-fold symmetry) over a half-length
of the tracker, and so each track finder $\phi$ nonant and half-length connects to $12$ DTCs (that is, 108 DTCs per half-length in $z$ split into nine in $\phi$). Within each $\phi$ nonant and half-length, the modules connected to each DTC are distributed in both the barrel and disk regions in order to better balance the stub load in each DTC, and to minimize or avoid multiple hits along a track being routed to a single DTC. A summary of the DTC layout within a $\phi$ nonant and half-length is shown in Table~\ref{tab:dtcs}.
We also assume a maximum of 72 modules connected per
DTC, and that each DTC handles a maximum output of \unit[672]{Gb/s}. This
system also requires that the sector processor boards are arranged
within two ATCA shelves for each $n^\mathrm{th}$ time-multiplexed slice.



\fixme{28 or 25? And certainly not 8b/10b...} 
As mentioned above, the DTCs will process the data from each of the modules in the tracker.  Each DTC will connect to one front-end cable, and each cable will handle 144 fibers; 72 front-end fibers running at \unit[10]{Gb/s} (inner detector layers) or \unit[5]{Gb/s} (outer detector layers) coming in, and 72 front-end fibers running at \unit[2.5]{Gb/s} going out. The 
DTC sends data to the track trigger using \unit[28]{Gb/s} cables, with a maximum capacity of \unit[672]{Gb/s} for each DTC (i.e., \unit[$24 \times 28$]{Gb/s} links).  The global coordinate information is sent to the track finder and uses \unit[36]{bits} per stub, and with \unit[64/66]{bit} encoding we assume a total of \unit[45]{bits} per stub.  



The layout of the DTCs' connections to the detector modules divides the detector into $\phi$-nonants,
and into two halves along $z$. Twelve DTCs are connected to each $\phi$-nonant and $z$ half, for a total
of $12 \times 18=216$ DTCs assuming a 9-sector processor configuration; this corresponds to one sector processor
per nonant. However, due to the curvature of tracks within the detector, a small amount of data
duplication is needed near the boundaries of sectors to ensure that track formation can take place
entirely within one sector processor. The tracker
is divided into 9 sectors at a critical radius $r_c$. 
Tracklets that are generated within a sector and, during the projection step, produce trajectories that cross $r_c$ within a particular sector, are kept in that sector. Tracklets that during the projection step instead cross $r_c$ in an adjacent sector are discarded (they will be
rediscovered by the adjacent sector). 
At radii above and below $r_c$, curves corresponding to the lowest
acceptable \pt are drawn from each edge of a sector, and all stubs that lie within these curves have
the potential to form tracks within the sector, and so must be sent to the corresponding sector processor.
The curves in adjacent sectors will overlap at radii above and below $r_c$, and it is in these small
regions where data duplication is necessary. The parameter $r_c$ is  tunable and can be chosen based on an
optimization of the algorithmic precision, algorithmic efficiency, load balancing in the DTCs, and
constraints on the number of links available to the DTCs. If the nine sector processors were aligned
with the DTC $\phi$-nonants, the overlap regions on either side of a sector would require each DTC
to communicate with three sector processors. To reduce the number of required output links from
the DTCs, the 9 sectors are offset from the $\phi$ nonants such that each of the DTC nonants only needs to communicate with at most two sector processors.


The cabling requirements are studied assuming 9 sectors, 18 time-multiplexed slices, and \unit[28]{Gb/s} links from the DTCs to the sector processors. These cabling studies have been performed using simulated \ttbar events, with an average of 200 overlaid pileup interactions, to determine whether, assuming the cabling configuration described above, \unit[28]{Gb/s} cables have sufficient bandwidth to deliver the stubs from the DTCs to the sector processors.  For a single \unit[28]{Gb/s} link, one can transfer information for a maximum of 210 stubs, and the proposed cabling configuration was found to be sufficient to transfer the stub information from the DTCs to the appropriate sector processors with zero stub loss.

The described cabling scheme requires 36 output links from the DTCs for
a time-multiplex factor of 18. This corresponds to 48 input links to each sector processor
(24 DTCs/nonant $\times$ 2 nonants) resulting in a total of 7776 total links
between the DTCs and sector processors. We conclude from these studies that the I/O
requirements of the algorithm are met by the system, and that \unit[28]{Gb/s} links
have enough bandwidth for transferring the stub information.



\begin{table}
  \caption{DTC layout in each nonant and half-length in $z$.  This layout
    yields a total of 216 DTCs. Each line corresponds to one DTC;
    there are 12 DTCs for each side of the detector ($+z$ and $-z$).}
  \label{tab:dtcs}
  \centering
  \begin{tabular}{cllll}
    \hline
	  Module type & \multicolumn{4}{l}{Connected detector regions} \\
    \hline
   PS & Layer 1 Flat Region & Disk 1, Rings 1-7 & Disk 3, Rings 1-3 & Disk 5, Rings 1-3 \\
         & Layer 1 Tilted Region & Disk 2, Rings 5-7 & Disk 4, Rings 1-3 & \\
         & Layer 2 & Disk 2, Rings 1-4 & & \\
         & Layer 3 Flat Region & Disk 2, Rings 8-10 & & \\
         & Layer 3 Tilted Region & Disk 4, Rings 4-7 & & \\
         & Disk 1, Rings 8-10 & Disk 3, Rings 4-7 & Disk 5, Rings 4-7 & \\
    \hline
    2S & Layer 4 & & & \\
         & Layer 5 & & & \\
         & Layer 6 & & & \\
         & Layer 6 & Disk 3, Rings 8-12 & & \\
         & Disk 1, Rings 11-15 & Disk 4, Rings 8-12 & & \\
         & Disk 2, Rings 11-15 & Disk 5, Rings 8-12 & & \\
    \hline
  \end{tabular}
\end{table}

%% file: performance.tex
\section{Physics performance}
\label{sec:performance}

In this section the expected performance of the tracklet algorithm is discussed. The studies use simulated data derived from a \textsc{Geant4}-based simulation of the CMS detector~\cite{geant4}. To evaluate the performance, a \textsc{C++} implementation of the algorithm is used. It models the limitations in the hardware implementation resulting from fixed-point arithmetic, finite buffer lengths, and similar constraints. This ``emulation'' algorithm, or emulator, truncates all calculations and bit widths as is done in the hardware, and implements all internal buffers and memories as they exist in the firmware. The firmware implementation, which runs on the hardware, is made to match the emulator bit-by-bit in our qualification process. 
The precision of the calculations used in the emulator is tuned to ensure sufficient accuracy to meet the required physics performance goals in the trigger.

The demonstrator setup for tracklet 1.0 shows excellent agreement between the actual firmware results and the integer-based \textsc{C++} emulation of the system. For single object events, the output tracks from the firmware have 100\% bitwise compatibility with the integer-based emulation. For busier events, for example \ttbar with an average pileup of 200, the emulation and firmware tracks agree to better than 99\%, as was shown in Figure~\ref{fig:sim_demo_comparison}. These tests validate the use of the integer-based emulation code for extrapolating to future performance improvements. This section shows the performance of the tracklet 2.0 configuration. 

%

\subsection{Performance of different steps of the tracklet algorithm}

\begin{figure}
\begin{center}
\includegraphics[width=0.7\textwidth]{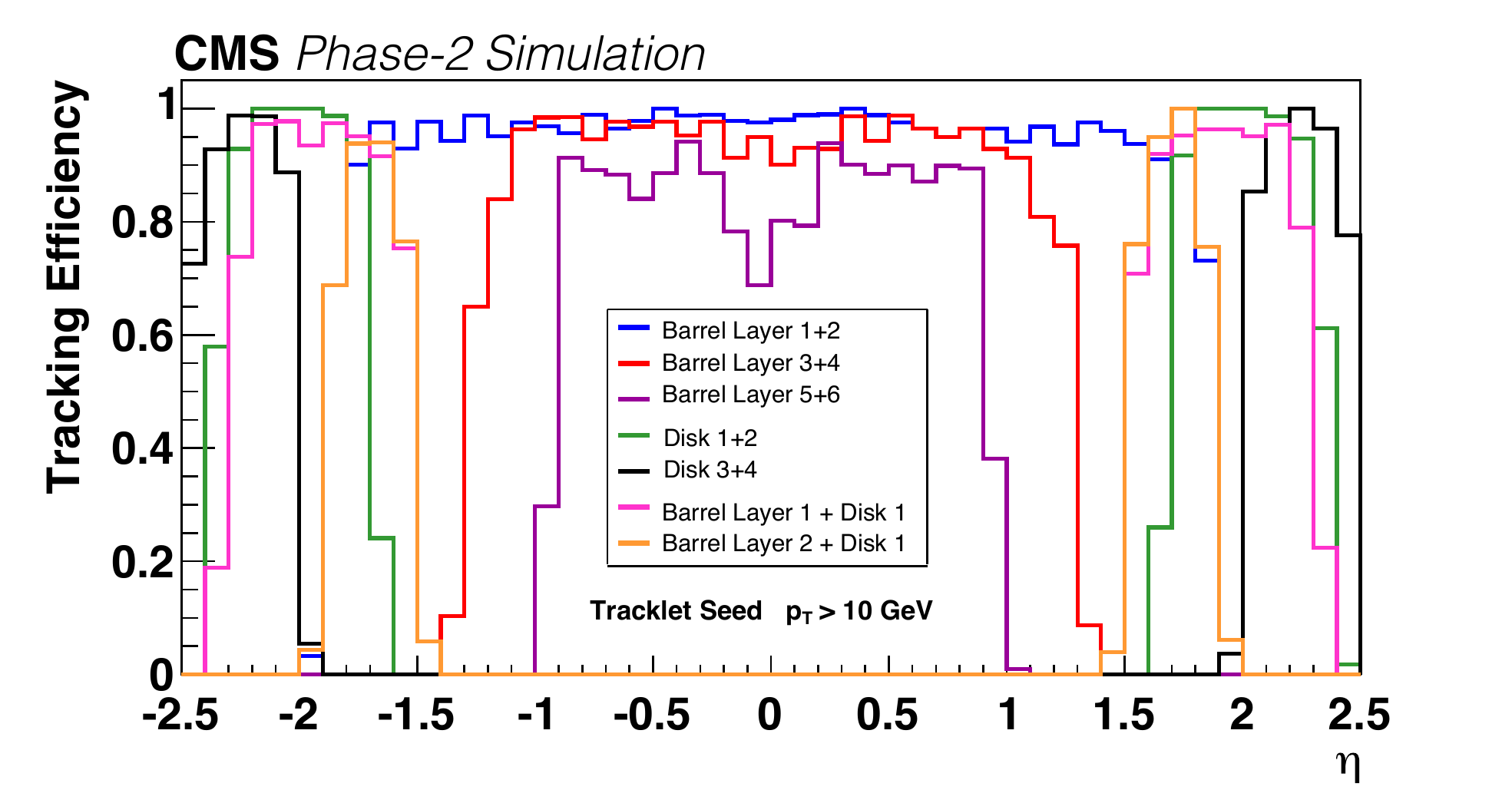}
\caption{\label{fig:seed_eff} Final tracking efficiency for single muons with $\pt > \unit[10]{GeV}$ in events without pileup, shown as a function of muon $\eta$ for different layer and disk combinations in forming the seed tracklets. 
The overlapping curves demonstrate the redundancy in seeding as a function of pseudorapidity. }
\end{center}
\end{figure}
\begin{figure}
\begin{center}
\includegraphics[width=0.49\textwidth]{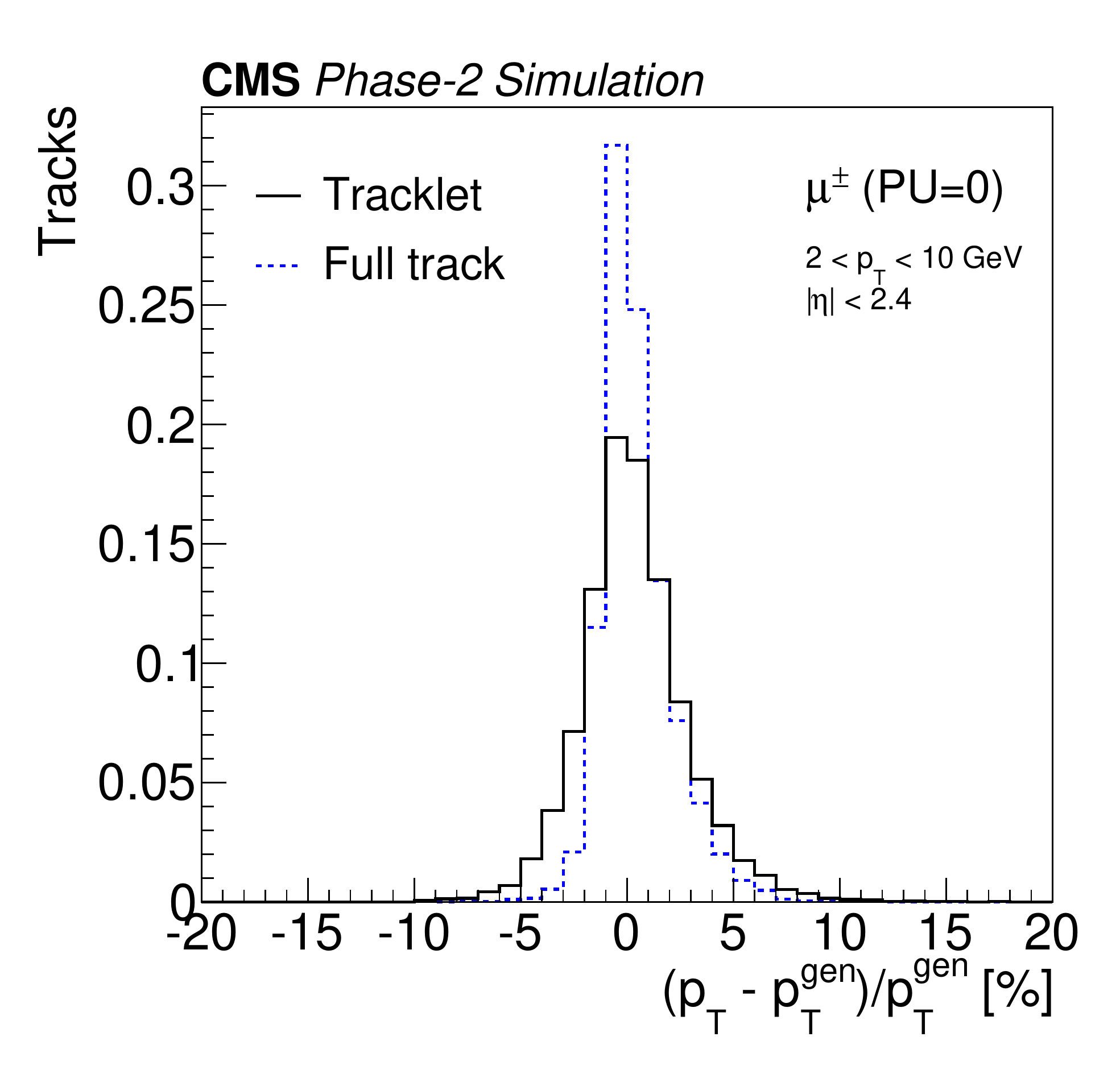}
\includegraphics[width=0.49\textwidth]{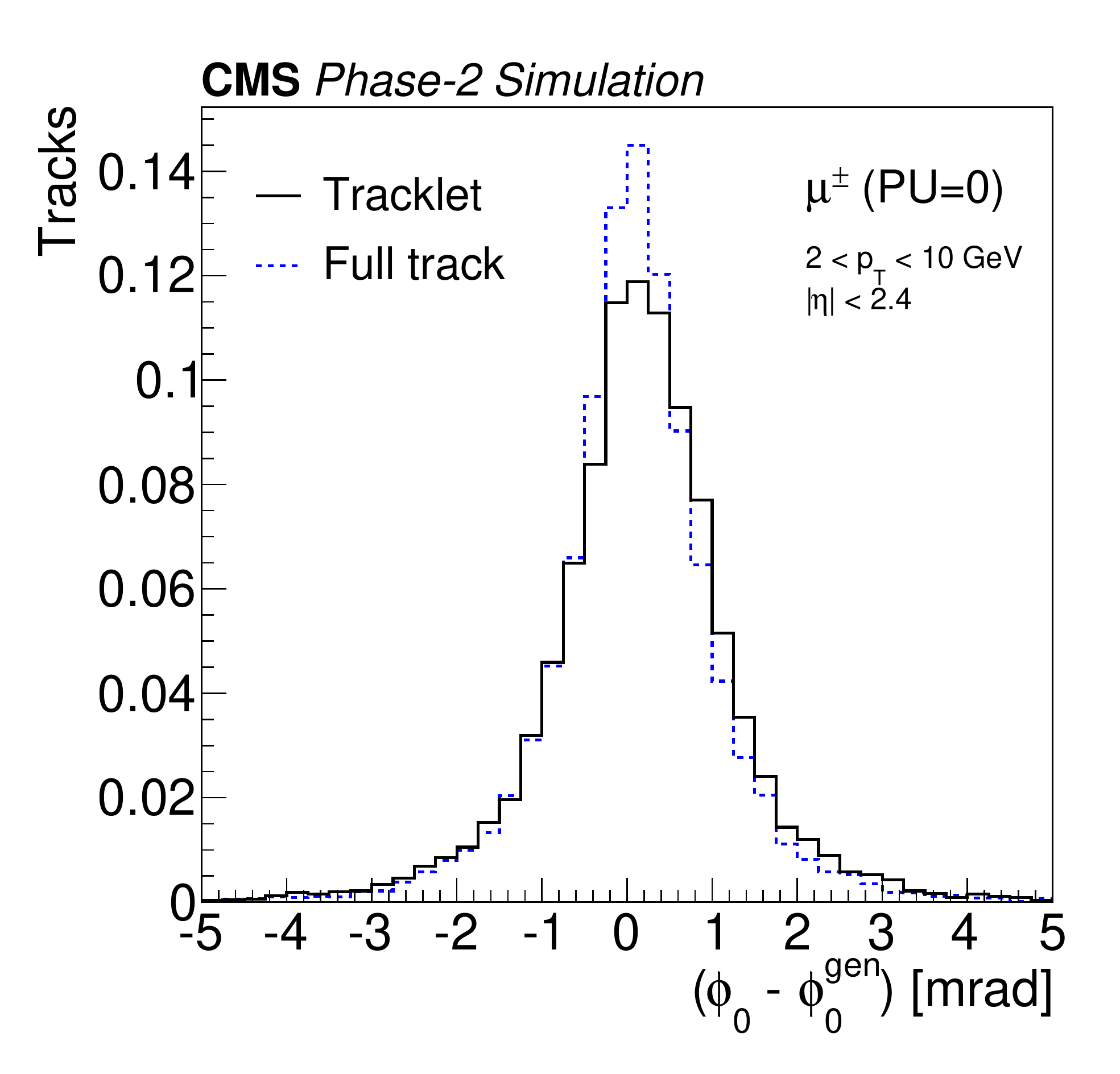} \\
\includegraphics[width=0.49\textwidth]{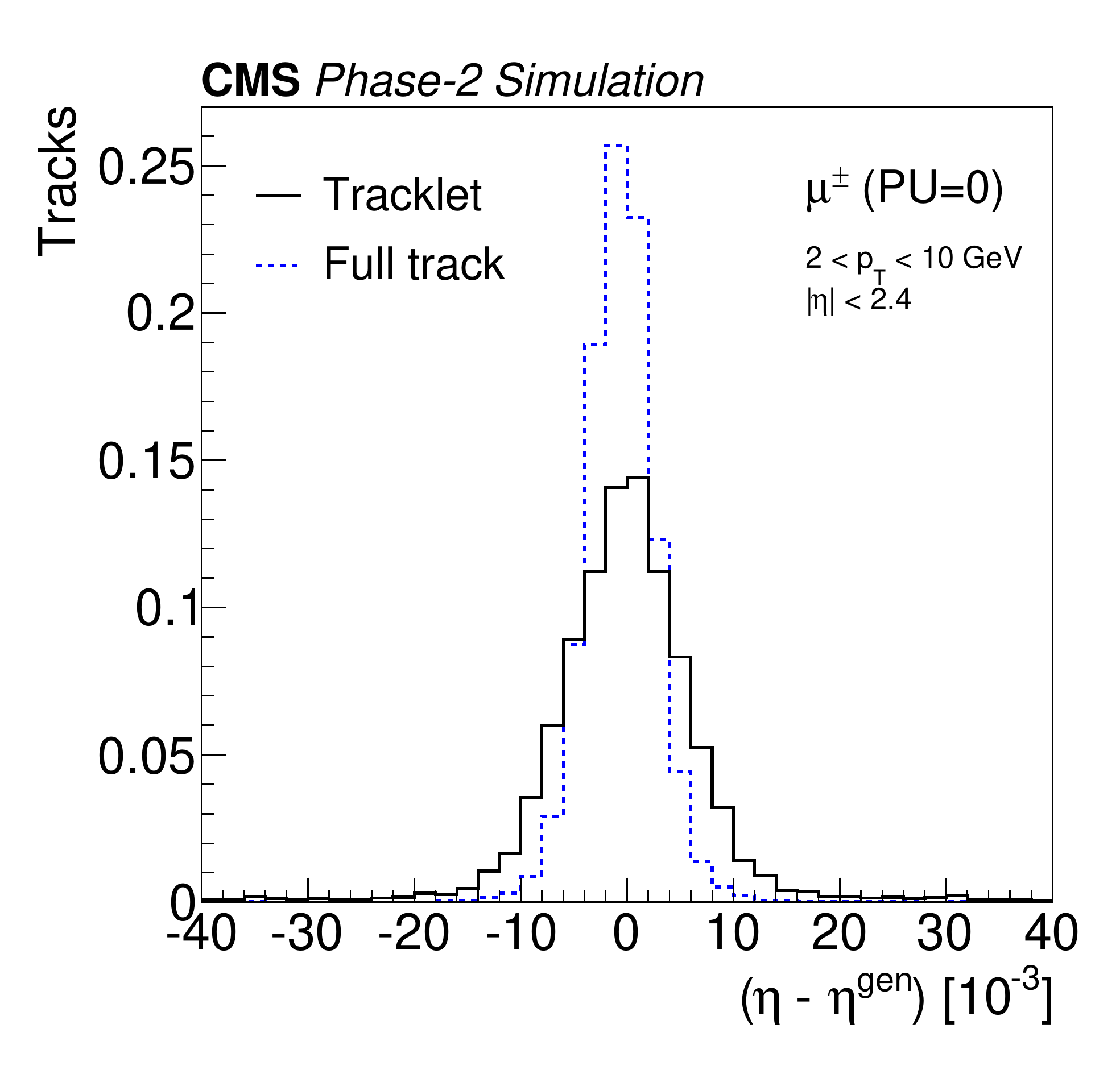}
\includegraphics[width=0.49\textwidth]{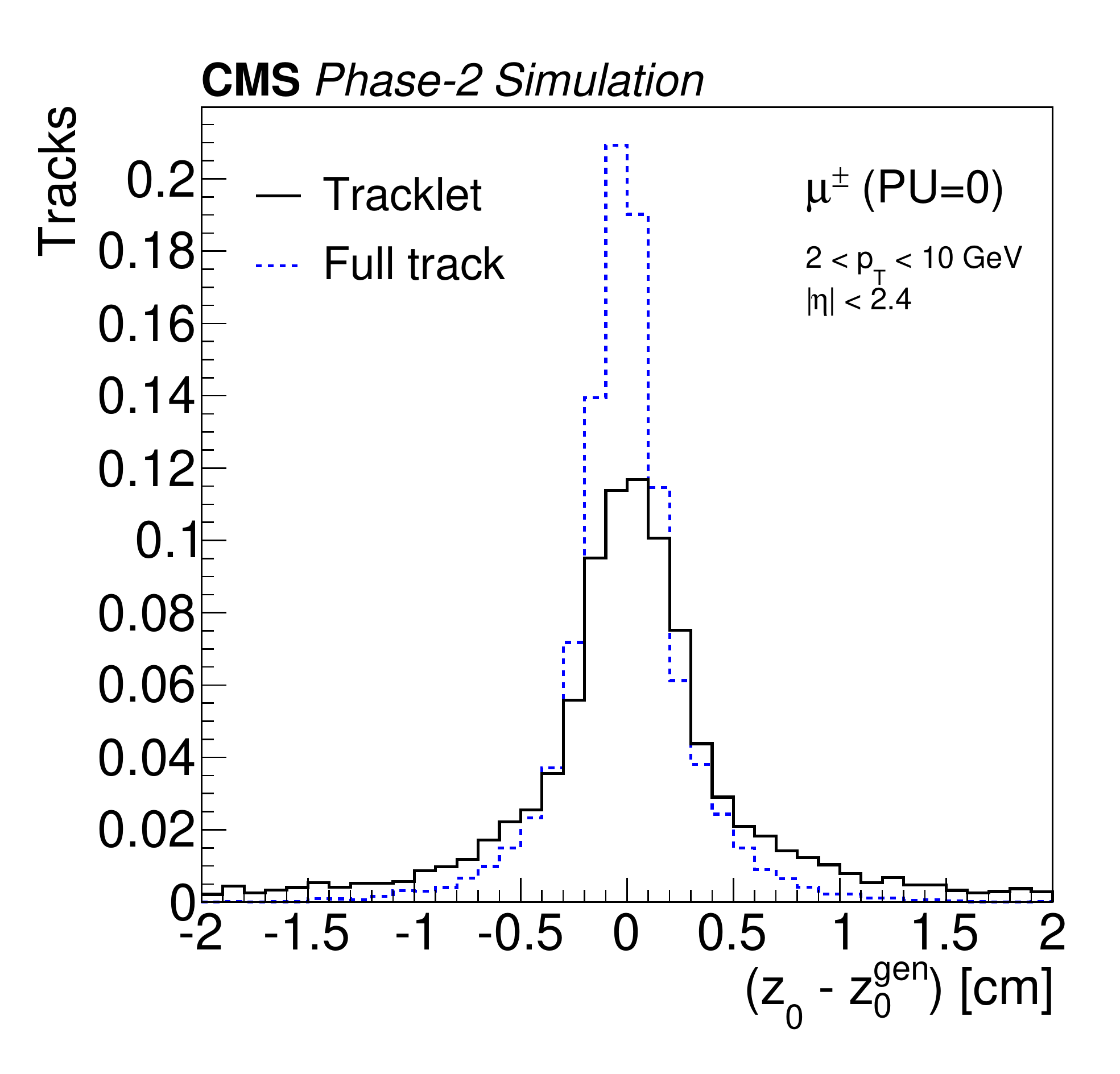}
\caption{\label{fig:tracklet_res} Residuals for relative $\pt$ (top left), $\phi_0$ (top right), $\eta$ (bottom left), and $z_0$ (bottom right) for the tracking seeds, tracklets, as solid black curves as compared to residuals for the full, final track fit in dashed blue curves. The track parameters are calculated using fixed-point calculations and shown for single muons ($\mu^{\pm}$) with $2< \pt < \unit[10]{GeV}$ and $|\eta|<2.4$ in events without pileup.}
\end{center}
\end{figure}

The efficiency for identifying charged particle trajectories, using different seeding combinations, is shown in Figure~\ref{fig:seed_eff} for a simulated sample of events. Each event contains a single muon without any additional pileup interactions, using the integer-based \textsc{C++} emulation of the algorithm. With the multiple seeding combinations, the same particle trajectory is often found multiple times, which ensures redundancy and a complete detector coverage.
As illustrated in Figure~\ref{fig:tracklet_res}, already the resolution of the tracking seeds (tracklets) is good -- within a factor of two or less of the resolution of the fitted track, motivating the feasibility of the road search algorithm. Given the resolution of the seeds, we can use narrow windows in projecting the tracklet seeds to other layers and disks to search for matching stubs. This is a key element in limiting the number of combinations that must be tried by the algorithm. 

The track parameters found by the TrackletCalculator are used to project
the tracklet to other layers and disks. This projection is performed in the 
TrackletCalculator to a nominal position of the layer or disk. In the 
MatchCalculator, once a candidate stub is found to which the projection is matched,
an exact projection is calculated using the derivatives
$\partial\phi_{\rm proj}/\partial r$ and $\partial z_{\rm proj}/\partial r$.
A projection-stub match is accepted if the $r$--$\phi$ and $z$ (or $r$ in disks) 
residuals pass a selection criterium that depends on (i) in which layer or 
disk the match is found and (ii) which combination of layers or disks were 
used to form the tracklet. 
The residuals are tuned to be nearly 100\% efficient for trajectories from prompt particles originating from the primary interaction vertex.

\begin{figure}
\begin{center}
\includegraphics[width=0.7\textwidth]{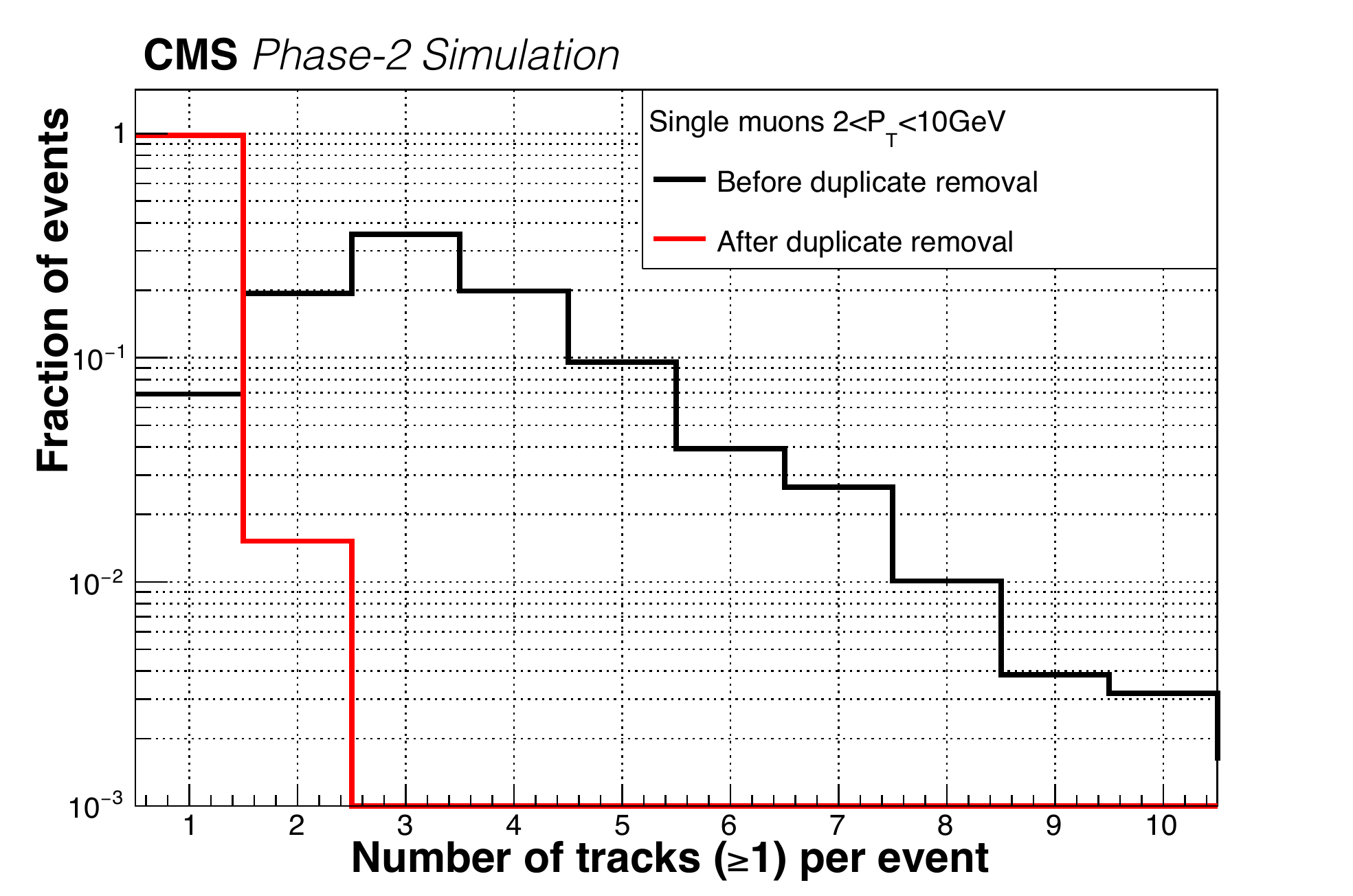}
\caption{\label{fig:duplicate_hourglass} Number of tracks per event for single muons with $2 < \pt < \unit[10]{GeV}$ in events without pileup. 
  The number of tracks is shown before any duplicate removal (black) and after duplicate removal within sectors (red). As expected, after duplicate removal, most events have exactly one track. Only events with at least one track prior to duplicate removal are shown.}
\end{center}
\end{figure}

Next, the performance of the duplicate removal is studied. Figure~\ref{fig:duplicate_hourglass} shows for single-muon events the number of tracks found per event prior to any duplicate removal and after the duplicate removal between tracks within a $\phi$ sector. For events containing a single muon, one track per event is expected, which is what is observed for most events after the duplicate removal step.

\subsection{Final tracking performance}

\begin{figure}
\begin{center}
\includegraphics[width=0.48\textwidth]{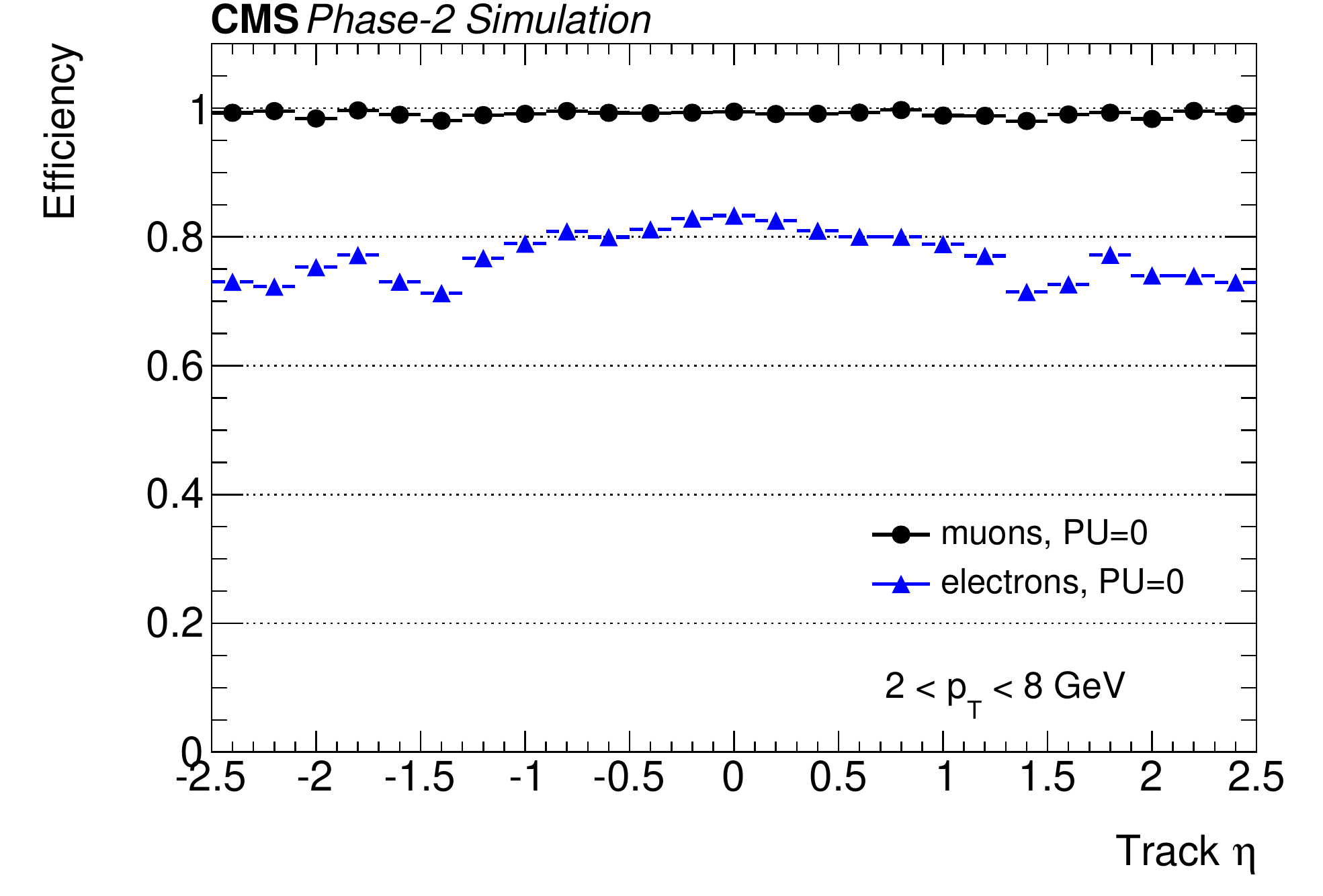}
\includegraphics[width=0.48\textwidth]{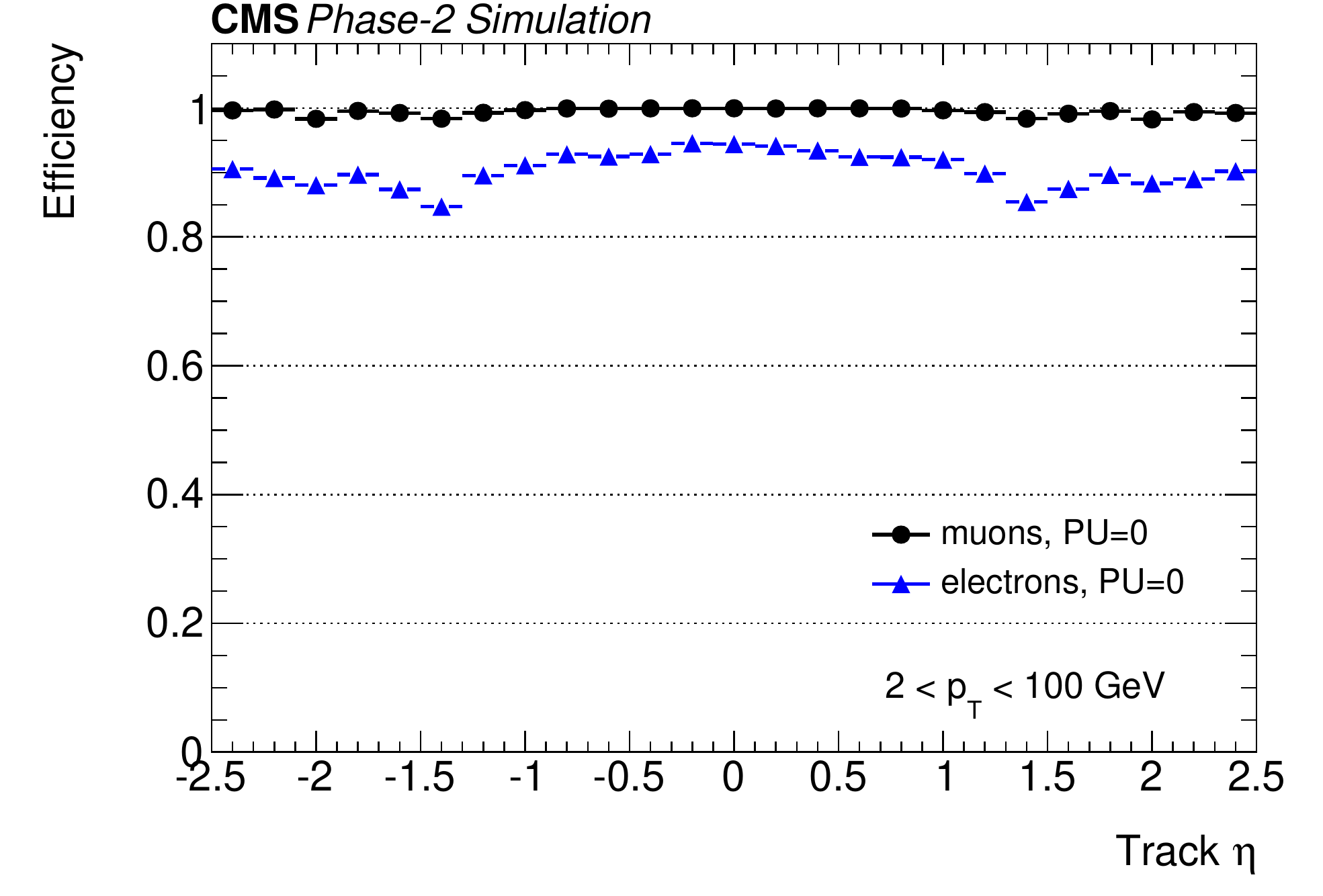} \\
\includegraphics[width=0.48\textwidth]{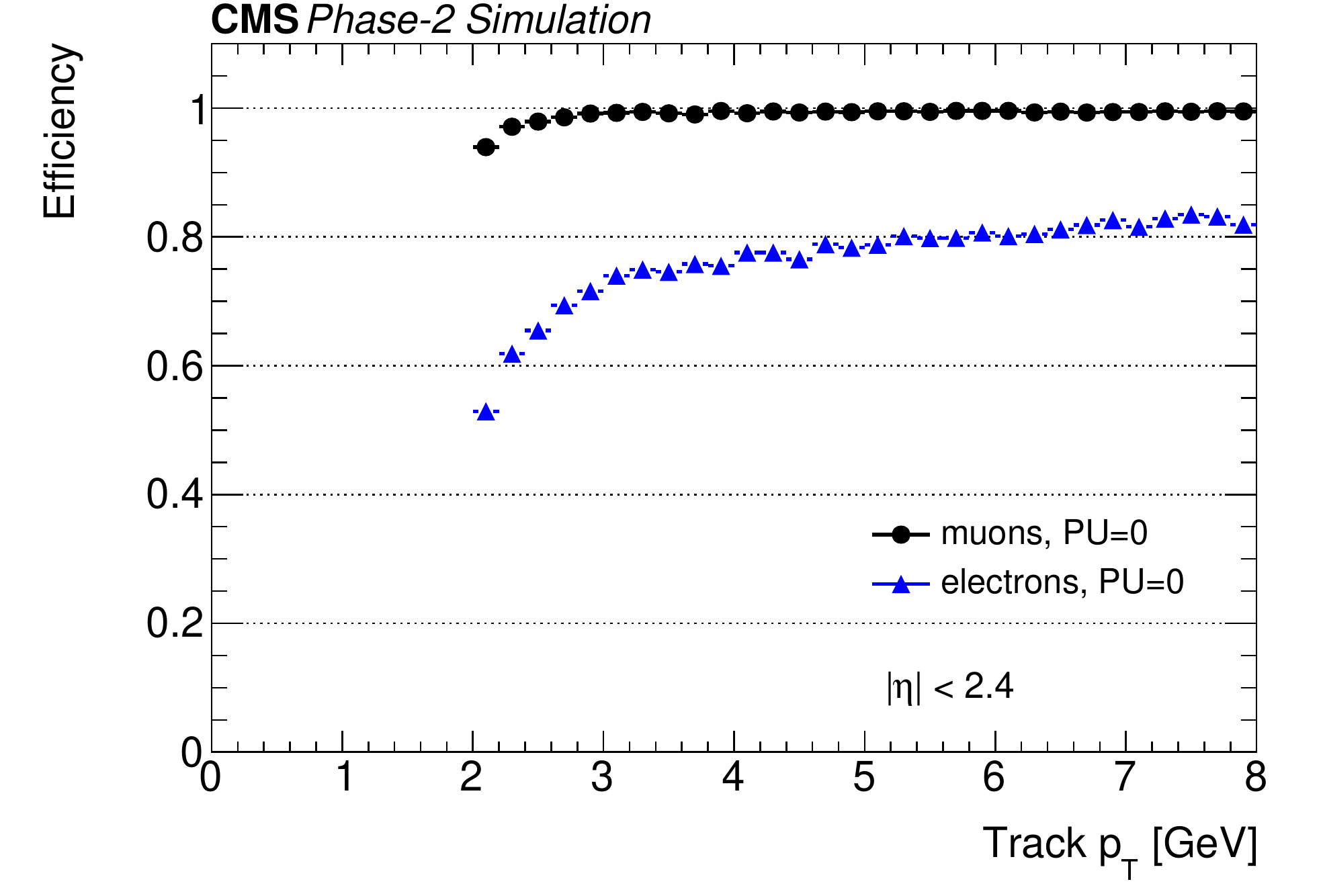}
\includegraphics[width=0.48\textwidth]{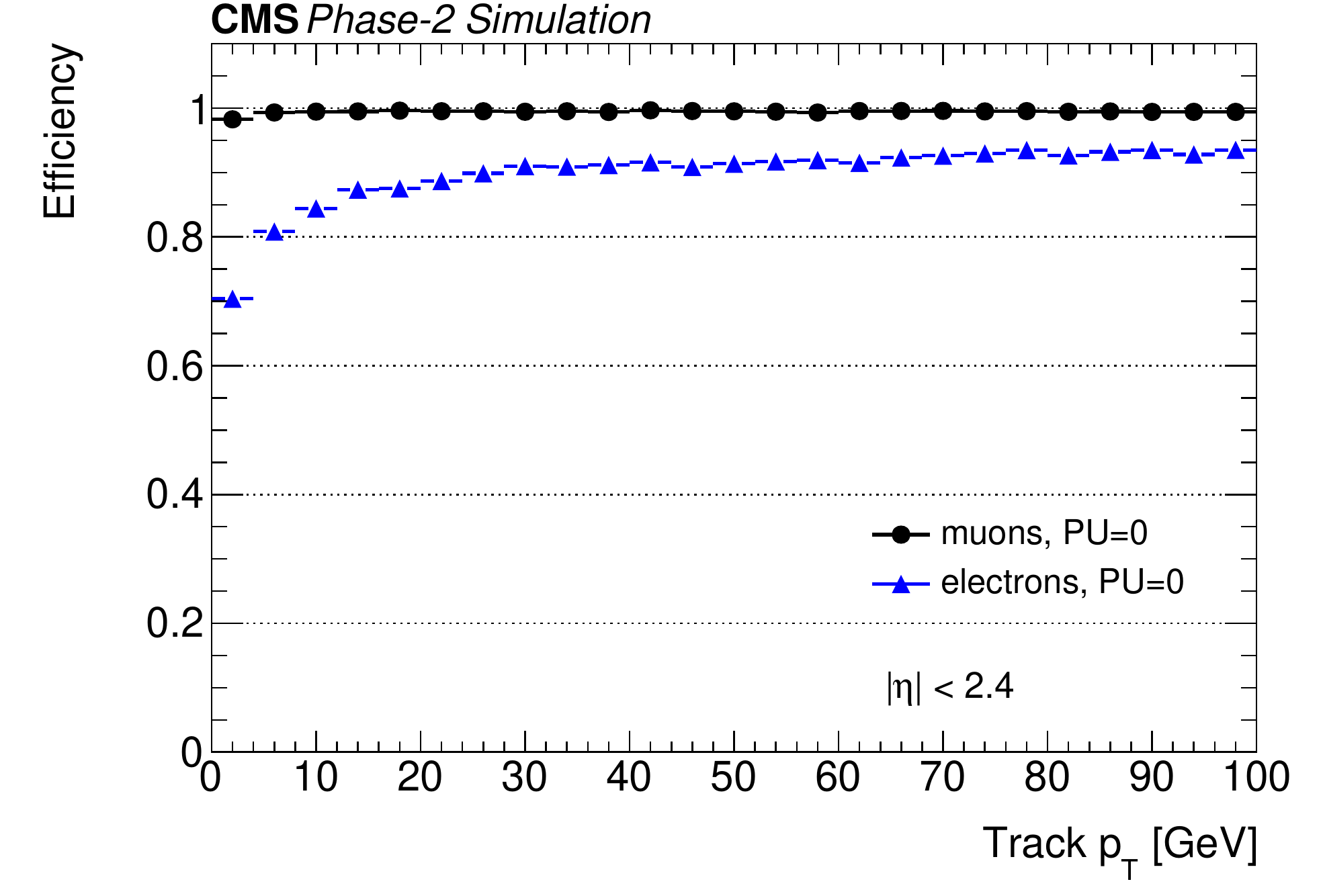}
\caption{\label{fig:emu_efficiencies} Efficiency as a function of $\eta$ (top row) and $\pt$ (bottom row) for muons and electrons with $2<\pt<\unit[8]{GeV}$ (left column) or $2<\pt<\unit[100]{GeV}$ (right column) in single-particle events without pileup. The propensity for electrons to radiate in the detector leads to a slower turn-on and lower plateau in the efficiency. }
\end{center}
\end{figure}

The estimated performance of the tracklet algorithm is studied with the integer-based emulation of the algorithm. The L1 tracking
efficiency as a function of $\pt$ and $\eta$ for single muons or electrons (generated with a uniform $\pt$ distribution) in events without pileup is shown in Figure~\ref{fig:emu_efficiencies}. Electrons have a lower efficiency compared to muons due to their higher interaction rate  with the detector material at the momenta in question. 

\begin{figure}
\begin{center}
\includegraphics[width=0.48\textwidth]{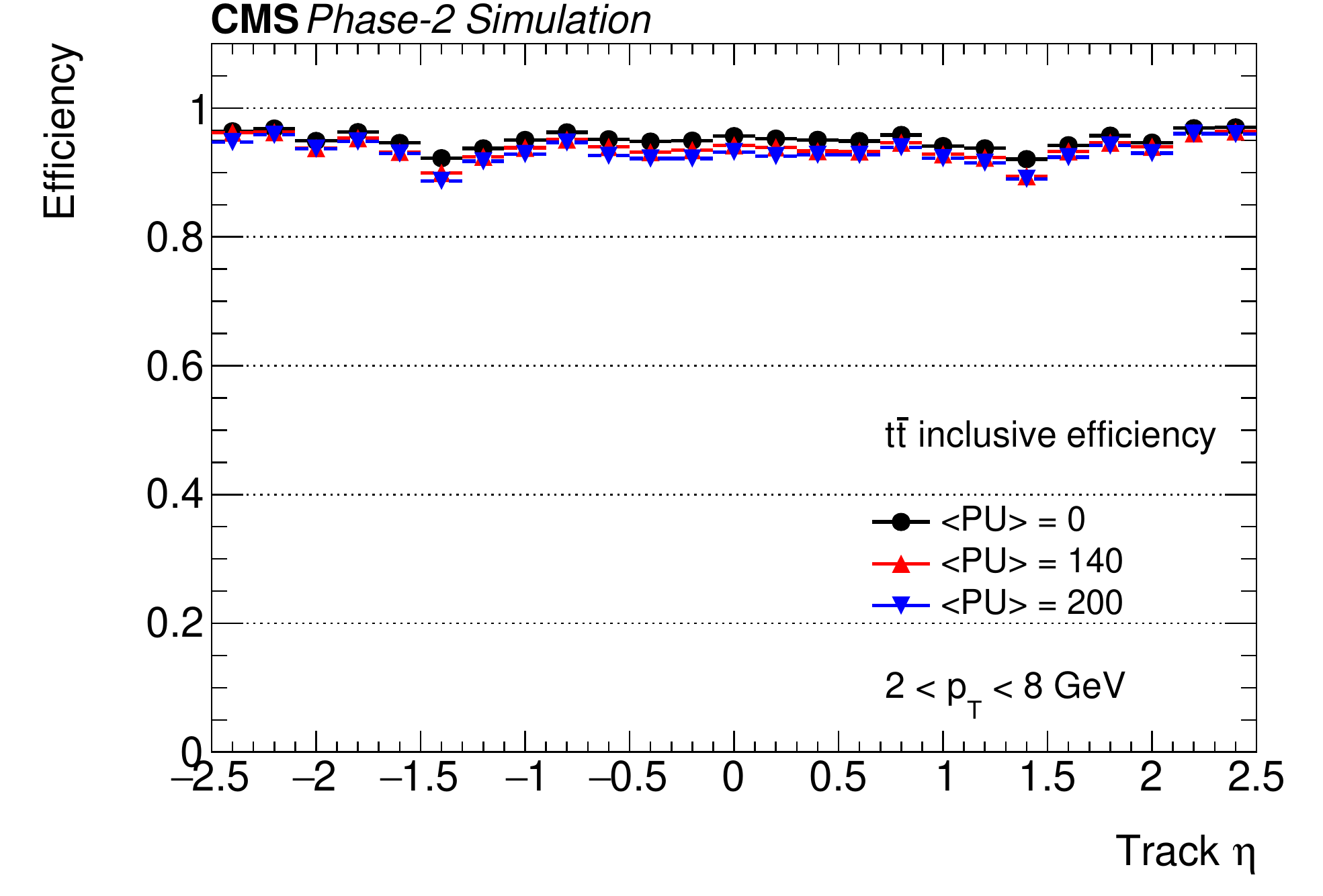}
\includegraphics[width=0.48\textwidth]{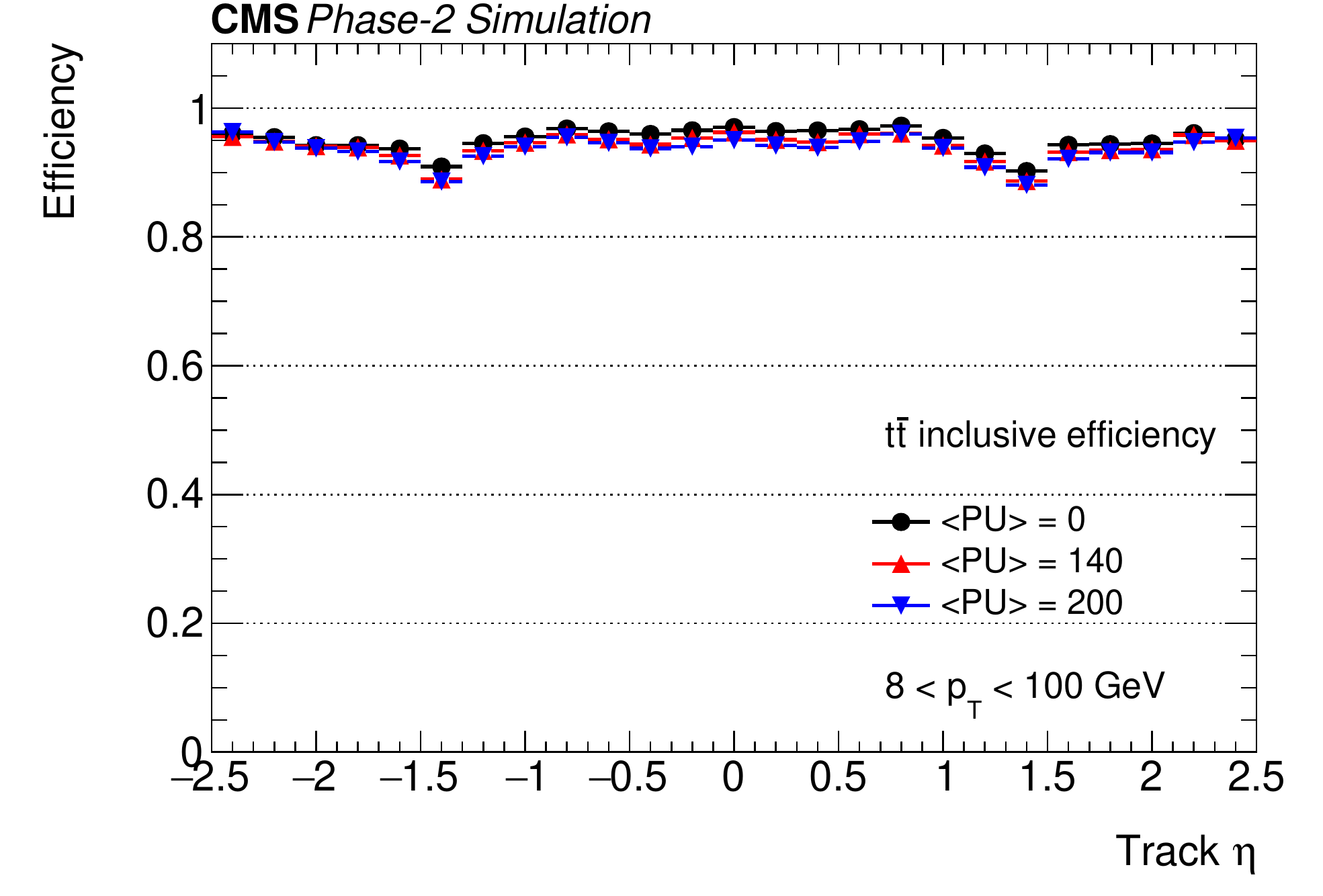} \\
\includegraphics[width=0.48\textwidth]{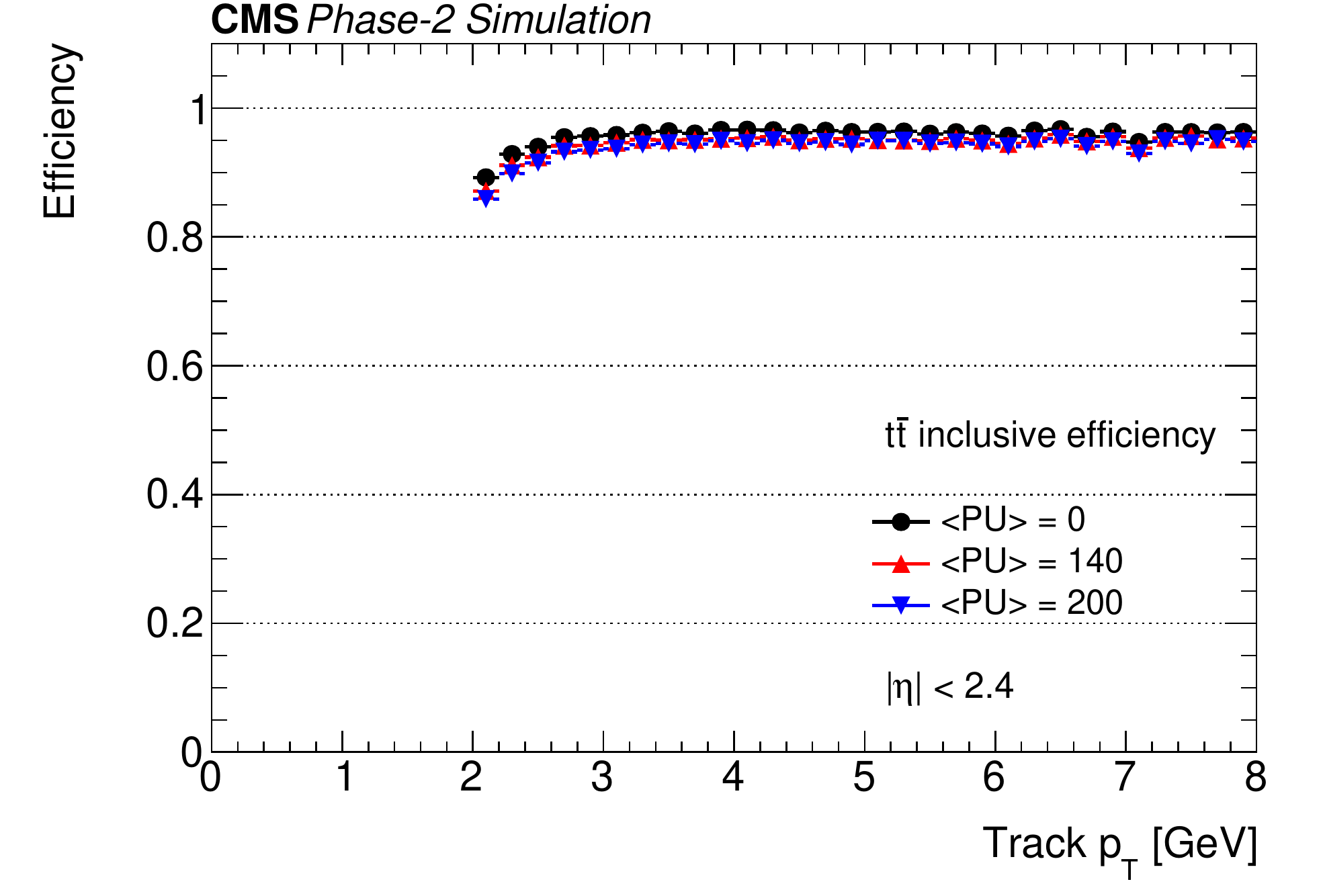}
\includegraphics[width=0.48\textwidth]{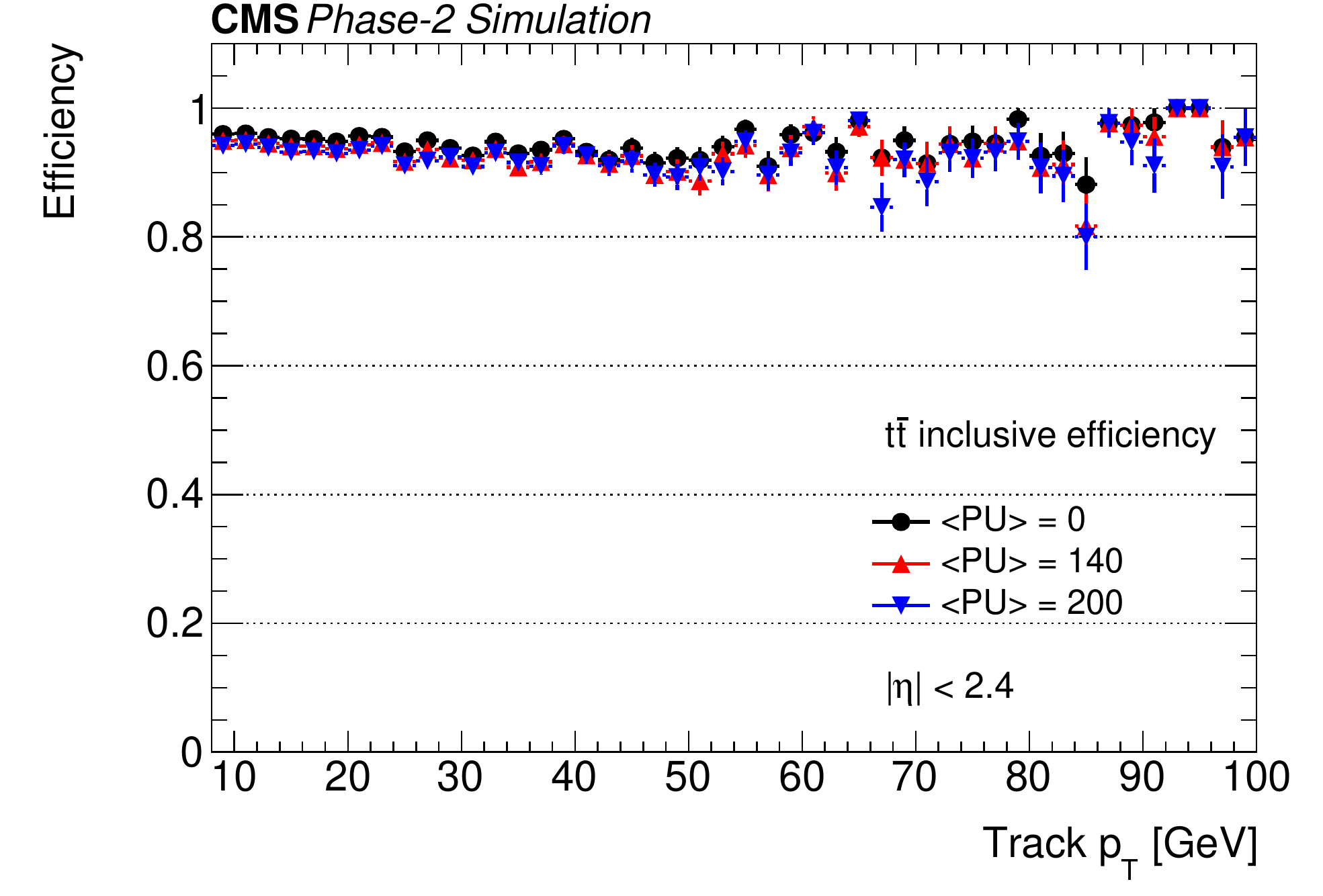} 
\caption{\label{fig:ttbar_efficiencies} Efficiency as a function of $\eta$ (top row) and $\pt$ (bottom row) for charged particles in \ttbar events, with an average pileup of 0, 140, or 200. Possible effects from truncation or the usage of fixed-point calculations are included.}
\end{center}
\end{figure}

Similarly in Figure~\ref{fig:ttbar_efficiencies}, the L1 tracking efficiency as a function of $\pt$ and $\eta$ is shown for charged particles from \ttbar production, in events with an average pileup of 0, 140, or 200. 
The efficiency is computed as implemented in the demonstrator, i.e., it includes the effects of truncating data. 
These effects are minimal, for two main reasons, (i) because of the large parallelization of the system, most of the modules are sparsely
populated, (ii) the different seeding combinations provide additional redundancy that can recover tracks that may otherwise be lost.

\begin{figure}
\begin{center}
\includegraphics[width=0.48\textwidth]{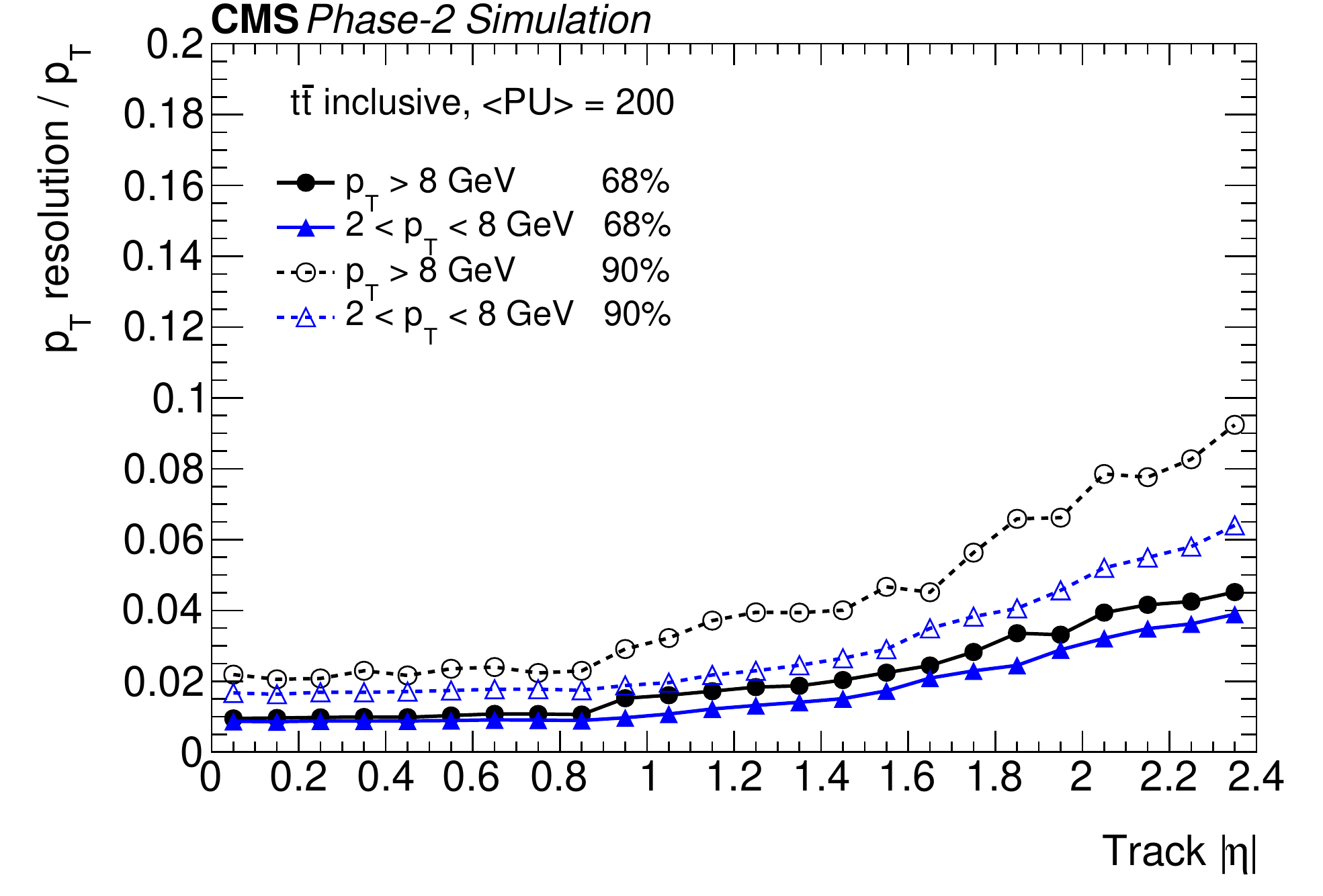}
\includegraphics[width=0.48\textwidth]{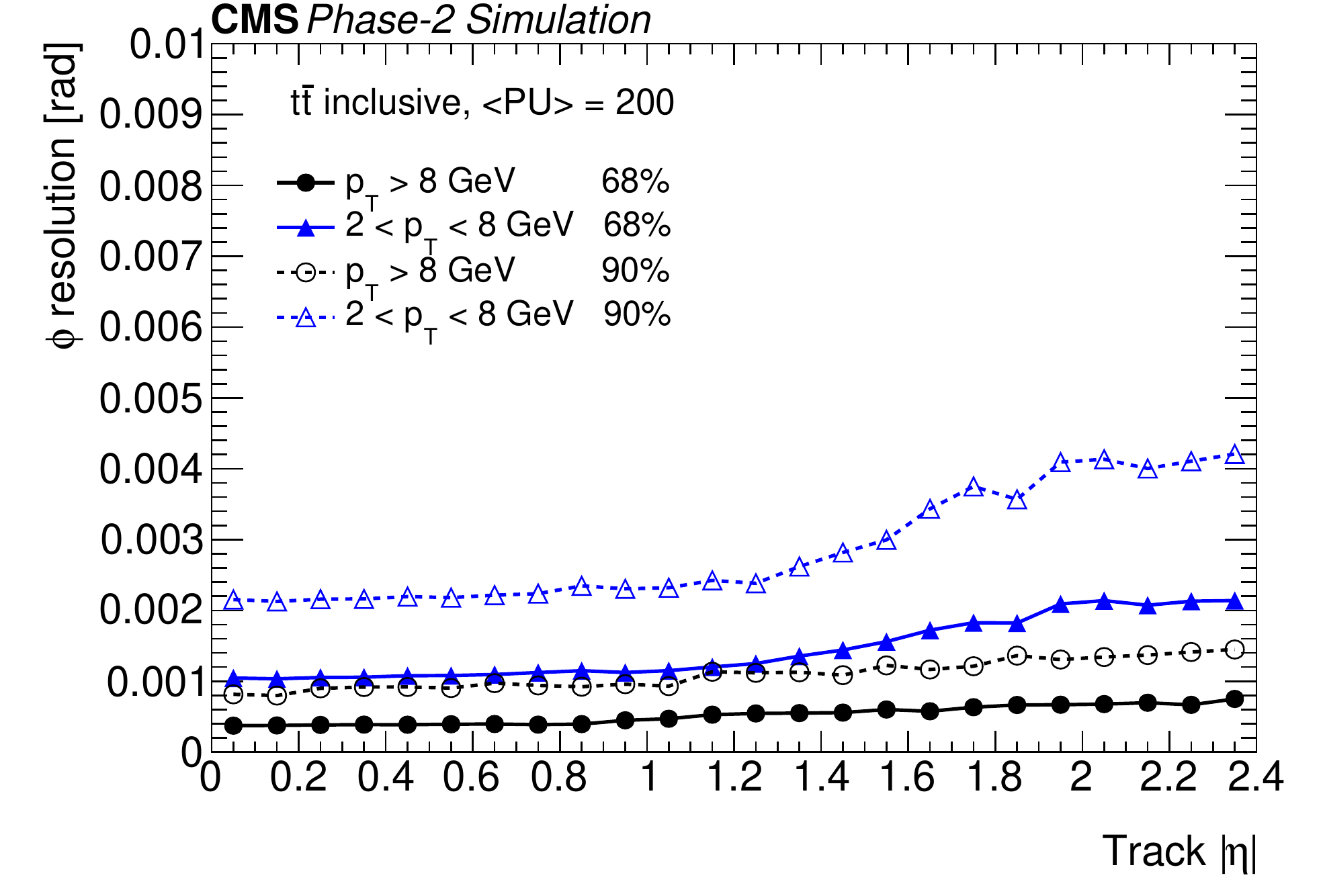} \\
\includegraphics[width=0.48\textwidth]{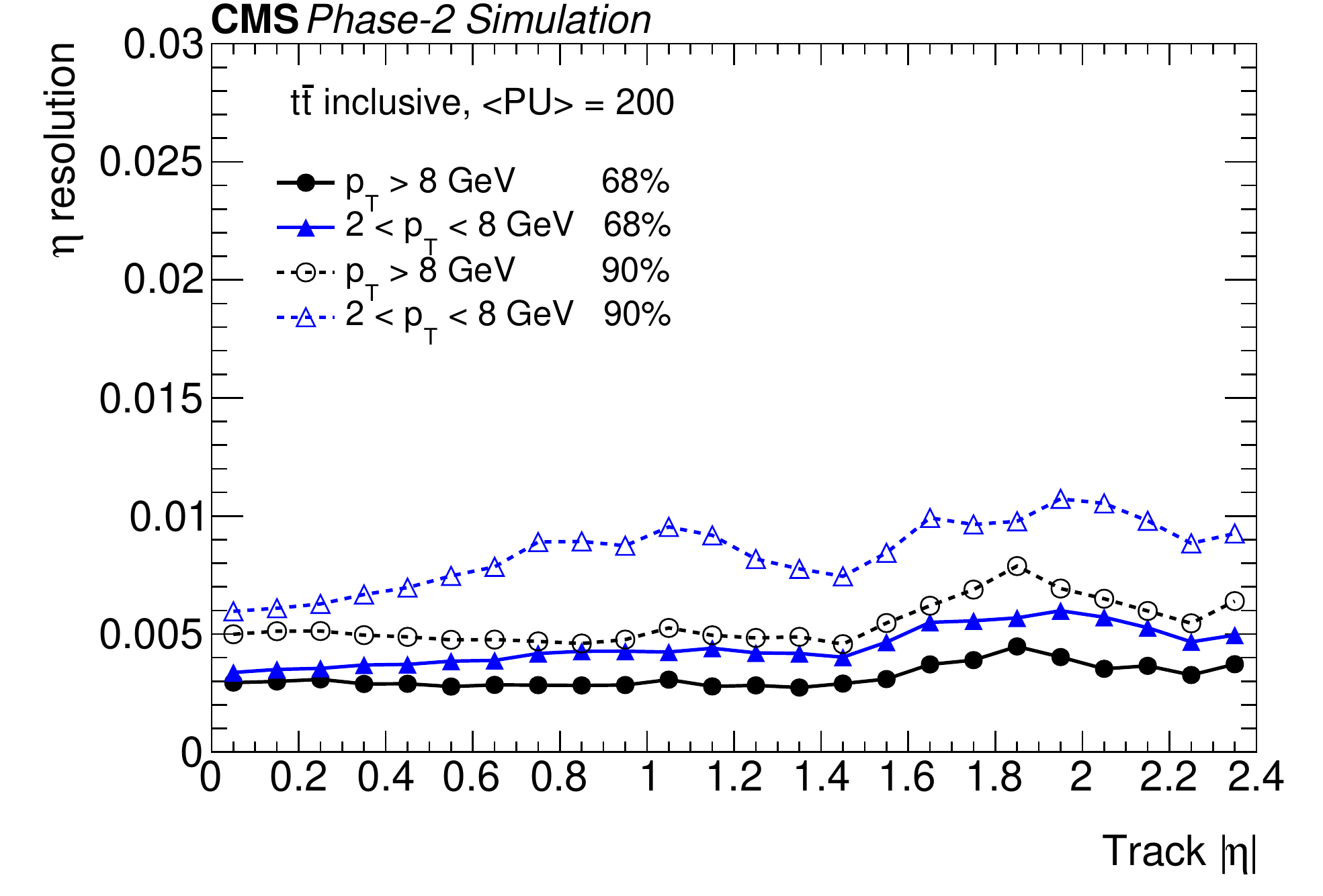}
\includegraphics[width=0.48\textwidth]{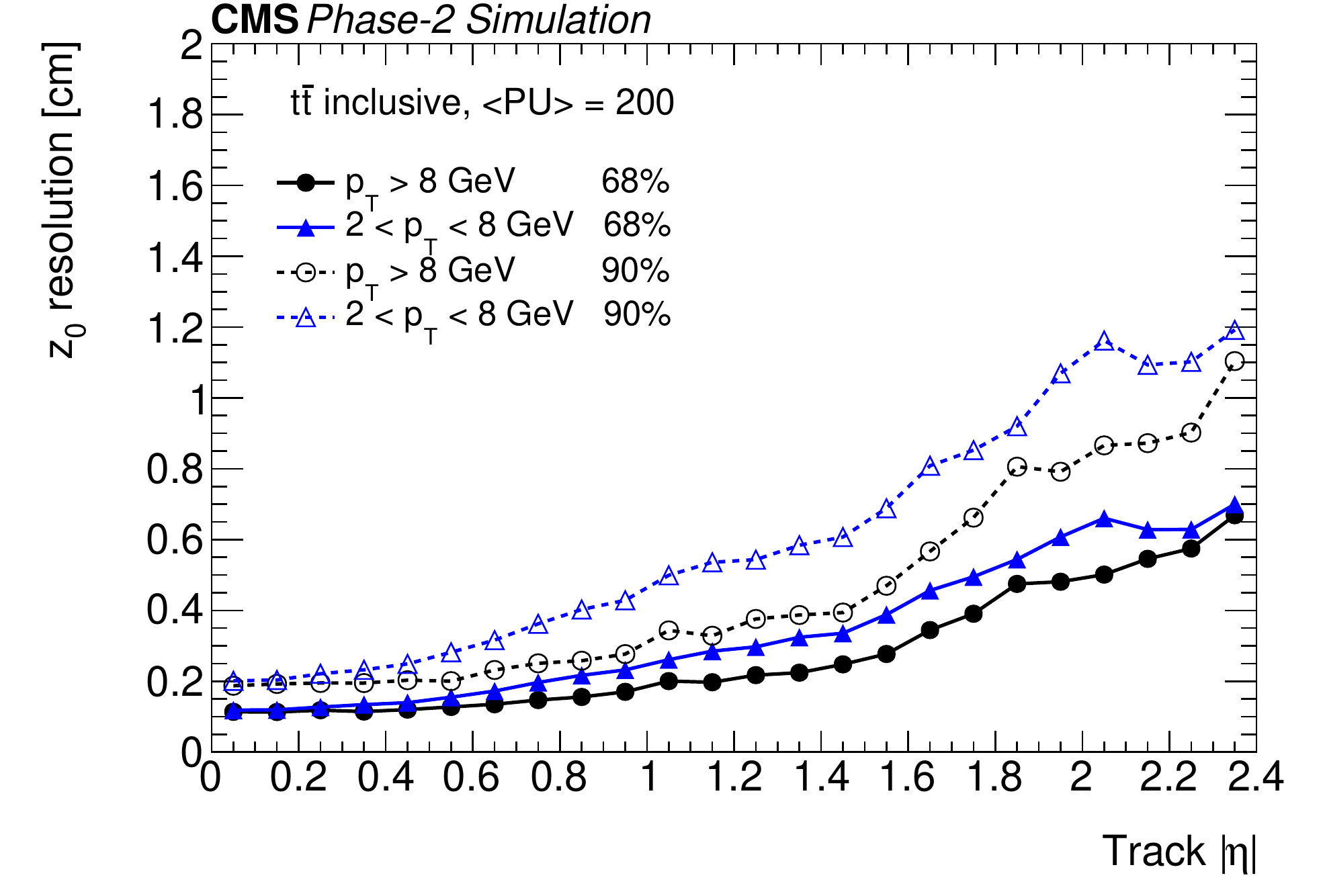} 
\caption{\label{fig:ttbar_resolutions} Relative $\pt$ resolution ($\sigma(\pt)/\pt$) (top left) and resolution in $\phi_0$ (top right), $\eta$ (bottom left), and $z_0$ (bottom right) for charged particles in \ttbar events with an average pileup of 200. The filled circles / solid lines correspond to the 68\% confidence intervals, while the open circles / dashed lines correspond to the 90\% confidence intervals. The resolutions are shown separately for tracks with $2<\pt<\unit[8]{GeV}$ and $\pt>\unit[8]{GeV}$. Possible effects from truncation or the usage of fixed-point calculations are included.}
\end{center}
\end{figure}

The L1 track parameter resolutions for $\phi$, $\eta$, $z_0$, and the relative resolution in $\pt$ are shown in Figure~\ref{fig:ttbar_resolutions} for charged particles from \ttbar production, in events with an average pileup of 200. The resolutions are shown as a function of $|\eta|$ for two different ranges in $\pt$ ($2<\pt<\unit[8]{GeV}$ and $\pt>\unit[8]{GeV}$). The $z_0$ resolution is about \unit[1]{mm} in the central barrel region, similar to the average separation of pileup vertices. The $z_0$ resolution  slightly worsens with increasing $|\eta|$ due to (i) the module tilts, and (ii), for $|\eta|>2.2$, the fact that a charged particle in this region does not cross a barrel layer with PS modules but only disk modules, which due to their orientation do not provide a precise $z$ measurement. 
The momentum resolution is about 1\% for central $\eta$ and increases to about 4\% for the outermost $\eta$ region. The momentum resolution is increased for higher values of $|\eta|$ due to the increased extrapolation distance to the beam axis, where the track parameters are calculated. 
The precise $z_0$ resolution allows the selection of tracks originating from a common vertex for use in L1 trigger algorithms and the accurate $\pt$ resolution results in sharp muon trigger thresholds.

\begin{figure}
\begin{center}
\includegraphics[width=0.6\textwidth]{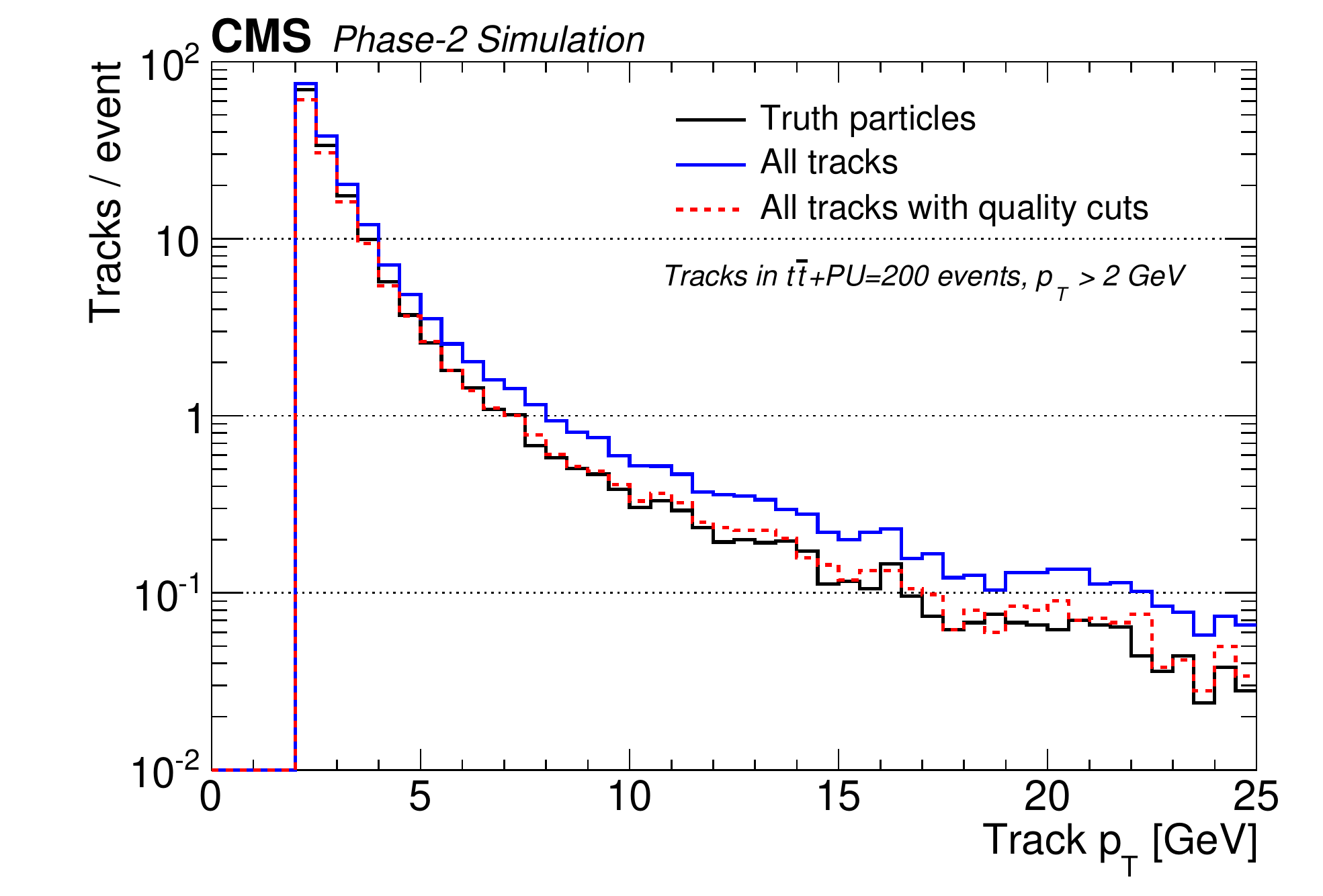} 
\caption{\label{fig:ttbar_rate} Track rate for $t\bar{t}$ events with an average of 200 pileup interactions. The lines show the truth track rate (solid black line), the rate of all reconstructed tracks above 2 GeV (solid blue line), as well as the rate of tracks after applying a set of quality criteria (dashed red line).}
\end{center}
\end{figure}

The track rates are shown in Figure~\ref{fig:ttbar_rate}. It shows the truth track rate, namely the rate of reconstructable particles that produce stubs in at least four layers or disks in the outer tracker (corresponding to the ``denominator'' in the efficiency plots). Also shown is the total rate of reconstructed L1 tracks above 2 GeV, as well as the rate of tracks when requiring quality criteria based on the stub bend information and the track fit $\chi^2$. The quality criteria reduces the contribution from misreconstructed (``fake'') tracks. These have, fractionally, a small contribution to the overall track rate, which is completely dominated by low-$\pt$ tracks where the fake rate is small. Minimizing the rate of fake tracks is regardless important for some use-cases of the L1 tracks, e.g. track-based missing transverse energy determination or hadronic tau lepton reconstruction. 

In summary, the track performance shown in this section achieves high track finding efficiency across the full $\pt$ and $\eta$ range covered. Additionally, track parameter resolutions are sufficiently precise for the downstream L1 trigger, where the tracks will either be correlated with information from other CMS detector systems, or used standalone to form tracker-only L1 trigger signatures.
